\documentclass[aps,prx,twocolumn,showpacs,superscriptaddress,longbibliography]{revtex4-2}
\usepackage{graphicx} 
\usepackage{amsmath,amssymb,graphicx}
\usepackage{graphicx,epstopdf}
\usepackage{gensymb}
\usepackage[dvipsnames]{xcolor}
\usepackage[T1]{fontenc}
\usepackage[utf8]{inputenc}
\usepackage{lipsum}
\usepackage{adjustbox}
\newlength\mylength
\newcommand{\av}{\mathbf a}
\newcommand{\bv}{\mathbf b}
\newcommand{\cv}{\mathbf c}

\newcommand{\be}{\begin{equation}}
\newcommand{\ee}{\end{equation}}
\newcommand{\bea}{\begin{eqnarray}}
\newcommand{\eea}{\end{eqnarray}}
\newcommand{\bse}{\begin{subequations}}
\newcommand{\ese}{\end{subequations}}

\usepackage{multibib}
\usepackage{color}
\usepackage[colorlinks,bookmarks=false,citecolor=darkblue,linkcolor=red,urlcolor=blue]{hyperref}

\definecolor{darkred}{rgb}{0.7,0.0,0.0}

\definecolor{darkblue}{rgb}{0,0.02,0.45}

\definecolor{darkgreen}{rgb}{0.02,0.45,0.0}

\definecolor{violet}{rgb}{0.8,0.2,0.6}

\begin{document}

\title{Proximate spin-liquid behavior in the double trillium lattice antiferromagnet K$_2$Co$_2$(SO$_4$)$_3$}
\author{A. Magar}
\author{K. Somesh}
\affiliation{School of Physics, Indian Institute of Science Education and Research Thiruvananthapuram-695551, India}
\author{M. P. Saravanan}
\affiliation{Low Temperature Laboratory, UGC-DAE Consortium for Scientific Research, University Campus, Khandwa Road, Indore 452001, India}
\author{J. Sichelschmidt}
\affiliation{Max Planck Institute for Chemical Physics of Solids, Nöthnitzer Strasse 40, 01187 Dresden, Germany}
\author{Y. Skourski}
\affiliation{Dresden High Magnetic Field Laboratory (HLD-EMFL), Helmholtz-Zentrum Dresden-Rossendorf, 01328 Dresden, Germany}
\author{M. T. F. Telling}
\affiliation{STFC Rutherford Appleton Laboratory, Didcot, Oxfordshire OX11 0QX, United Kingdom}
\author{V. A. Ginga}
\author{A. A. Tsirlin}
\affiliation{Felix Bloch Institute for Solid-State Physics, Leipzig University, 04103 Leipzig, Germany}
\author{R. Nath} 
\email{rnath@iisertvm.ac.in}
\affiliation{School of Physics, Indian Institute of Science Education and Research Thiruvananthapuram-695551, India}

\begin{abstract}
We report proximate quantum spin liquid behavior in K$_2$Co$_2$(SO$_4$)$_3$ with the magnetic Co$^{2+}$ ions embedded on a highly frustrated three-dimensional double trillium lattice. Single-crystal and high-resolution synchrotron powder x-ray diffraction experiments reveal a structural phase transition at $T_{\rm t} \simeq 125$~K from high-temperature cubic to low-temperature monoclinic phase with the three-fold superstructure. Magnetization and heat capacity consistently show the formation of the $J_{\rm eff} =1/2$ state of Co$^{2+}$ below 50\,K. In zero field, K$_2$Co$_2$(SO$_4$)$_3$ shows signatures of static magnetic order formed below $T^* \simeq 0.6$~K, but muon spin relaxation experiments reveal a large fluctuating component that persists down to at least 50\,mK, reminiscent of quantum spin liquid (QSL). Static order is completely suppressed in the small magnetic field of $\sim 1$\,T, and low-temperature heat capacity demonstrates the $T^2$ behavior above this field, another fingerprint of QSL. \textit{Ab initio} calculations show a competition of several antiferromagnetic couplings that render K$_2$Co$_2$(SO$_4$)$_3$ a promising pseudospin-$\frac12$ material for studying quantum magnetism in the double trillium lattice geometry.
\end{abstract}
\maketitle

Quantum fluctuations are a key ingredient for the exotic behavior of frustrated magnets. Ultimately, they can destabilize magnetic order and trigger the formation of a quantum spin liquid (QSL), a long-range-entangled state with highly nontrivial properties and fractionalized excitations. The QSL is typically envisaged in two-dimensional (2D) magnets with the triangular, kagome, and more complex geometries~\cite{Broholm6475,chamorro2021}, although some of the 3D geometries can be suitable for the QSL formation too~\cite{Chillal2020}.

From the experimental side, one promising group of 3D frustrated magnets are transition-metal compounds with the so-called trillium lattice formed by a corner sharing of three triangular units. This geometry is realized in the langbeinite family with the general formula $A_2B_2$(SO$_4$)$_3$ where $A$ and $B$ are the monovalent and divalent cations, respectively. 
Compounds of this family have been of interest because of their absence of inversion symmetry (space group $P2_13$) and ferroelectric as well ferroelastic transitions on cooling~\cite{Hikita1327,Percival563}. 

Recently, magnetic properties of K$_2$Ni$_2$(SO$_4$)$_3$ (abbreviated as NiKS) have been rigorously investigated in view of the field-induced QSL~\cite{Ivica157204}. NiKS has two structurally distinct Ni$^{2+}$ ($S=1$) ions that form two independent trillium lattices (double-trillium geometry). Despite a peak in the heat capacity data at $T_{\rm N}\sim 1.1$~K, the muon spin relaxation ($\mu$SR) rate reveals persistent dynamics down to low temperatures with only a small static component~\cite{Ivica157204}. In a magnetic field of $\mu_0H \geq 4$~T, the ordered component is fully suppressed, resulting in a field-induced QSL state. The excitation continuum revealed by inelastic neutron scattering (INS) further demonstrates the proximity to a QSL state~\cite{Yao146701}. Subsequent theoretical studies elucidated this proximity in terms of two leading magnetic interactions in the double-trillium geometry, $J_4$ and $J_5$, both antiferromagnetic~\cite{Yao146701,Gonzalez7191}. Additionally, several high-spin $S=3/2$ and $5/2$ compounds with the double-trillium geometry and a mixture of monovalent and divalent $A$-cations were recently investigated, and exotic field-induced features were reported~\cite{Boya101103,Khatua184432,Kolay224405}.



Reducing the local magnetic moment enhances quantum fluctuations and stabilizes the QSL. Therefore, a spin-$\frac12$ material with the trillium-lattice geometry would be of particular interest for exploring the QSL physics in 3D, but no such material has been investigated to date. Here, we consider a hitherto overlooked member of the langbeinite family, K$_2$Co$_2$(SO$_4$)$_3$ abbreviated as CoKS in the following. 
Several previous reports used dielectric, heat capacity, and optical absorption spectroscopy to detect the low-temperature structural transition in CoKS and related langbeinite materials~\cite{Hikita1327,Brezina623,Percival563}, but the nature of the transition and the frustrated magnetic behavior of CoKS were not investigated.

In this Letter, we report a comprehensive study of CoKS and identify this material as the first spin-$\frac12$ magnet with the trillium-lattice geometry. Our experiments establish the pseudospin-$\frac12$ nature of Co$^{2+}$ magnetism and reveal the proximate QSL state with the residual magnetic transition $T^* \sim 0.6$~K in zero field. Despite this transition, $\mu SR$ relaxation data pinpoint highly fluctuating spins with a small static component down to 50~mK. Concurrently, our \textit{ab initio} calculations reveal several competing antiferromagnetic interactions that place CoKS in the regime where QSL behavior can be expected theoretically.  We also resolve the structural phase transition in CoKS and demonstrate its minimal impact on the frustrated magnetic behavior.


The details of single-crystal growth and experimental characterization, as well as \textit{ab initio} calculations are 
given in the supplemental material (SM)~\cite{supplementary}. Since the nature of the low-temperature phase forming in CoKS below its structural phase transition temperature ($T_t$) has not been resolved in the previous literature, we performed single-crystal x-ray diffraction (SC-XRD) measurements at room temperature (295~K) and 90~K. 
The refinement of the room-temperature data confirms the cubic structure with the space group $P2_13$. At 90\,K, superstructure reflections indicate a three-fold supercell with the monoclinic symmetry (space group $P2_1$). 

\begin{figure}
\includegraphics[width=\columnwidth]{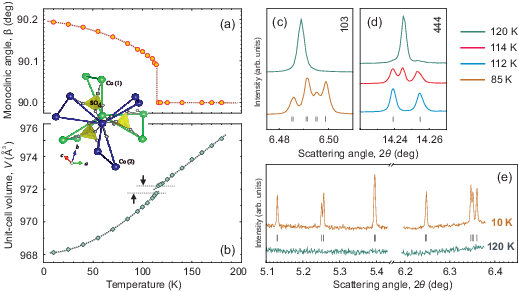}
\caption{\label{Fig1}
Temperature dependence of the (a) monoclinic angle $\beta$ and (b) unit-cell volume $V$,
both given for the subcell of the CoKS structure. Splitting of the (c) $103$ and (d) $444$ reflections below $T_t$ indicates monoclinic symmetry of the low-$T$ phase. The data collected at 114\,K additionally show phase coexistence around $T_t$. (e) Weak reflections appearing below $T_t$ manifest the three-fold superstructure. Tick marks indicate the Bragg peak positions for the low-$T$ phase. Dotted lines are guide-for-the-eye only. Note that the transition temperature of 114\,K in this experiment is somewhat lower compared to $T_t\simeq 125$\,K determined by thermodynamic measurements. This is due to the heating of the sample by the intense x-ray beam. The inset in (a,b) shows the frustrated geometry of CoKS with two interpenetrating trillium lattices built by two distinct Co sites.}
\end{figure}

Temperature-dependent lab XRD measurement fails to resolve any signatures of this structural phase transition (see SM~\cite{supplementary}), similar to Ref.~\cite{Moriyoshi4726}. Therefore, we used high-resolution synchrotron XRD to reveal the peak splitting below $T_t$. The low-symmetry structures of langbeinite compounds are usually orthorhombic ($P2_12_12_1$) or monoclinic ($P2_1$). In our case, the splitting of the $hhh$ reflections into two [Fig.~\ref{Fig1}(d)] serves as a direct fingerprint of the monoclinic symmetry. Additionally, a series of weak reflections appearing below $T_t$ manifest the three-fold superstructure of the low-temperature phase [Fig.~\ref{Fig1}(e)]. Its lattice translations are obtained as $\av_{\rm LT}=\av+\cv$, $\bv_{\rm LT}=\bv$, $\cv_{\rm LT}=\av-2\cv$ where $\av$, $\bv$, and $\cv$ are the lattice vectors of the high-temperature cubic phase. 

Temperature dependence of the lattice parameters extracted from the synchrotron XRD data reveals that the transition at $T_t$ is of first order. The monoclinic angle $\beta$ decreases upon heating but shows a discontinuous change at $T_t$ [Fig.~\ref{Fig1}(a)]. Likewise, the unit-cell volume drops by about 0.1\% at the transition [Fig.~\ref{Fig1}(b)]. Finally, the data collected near $T_t$ revealed a coexistence of the low-temperature monoclinic and high-temperature cubic phases manifested by three separate peaks at the positions of the $hhh$ reflections where only two peaks would be expected in the monoclinic phase [Fig.~\ref{Fig1}(d)]. The first-order nature of the transition implies that the distortion involves a combination of several modes.

Both the high-temperature cubic and low-temperature monoclinic structures feature disconnected CoO$_6$ octahedra that are linked through the SO$_4$ tetrahedra into a 3D framework with the double trillium lattice geometry [inset of Fig.~\ref{Fig1}(a)].
The main difference between the two structures lies in the additional cooperative rotations of the octahedra and tetrahedra in the low-temperature phase. The CoO$_6$ octahedra are weakly deformed already in the cubic phase with the Co--O distances of $2.05-2.10$\,\r A. These deformations are only slightly enhanced in the monoclinic phase where the Co--O distances lie in the range of $2.04-2.13$\,\r A. The transition is mainly driven by the K$^+$ ions that are underbonded in the cubic structure with the average bond valence~\cite{brown1985} of 0.79, and achieve the higher average valence of 0.96 in the monoclinic structure. Full structural parameters of the monoclinic phase are given in the SM~\cite{supplementary}.


\begin{figure}
	\includegraphics[width=\columnwidth]{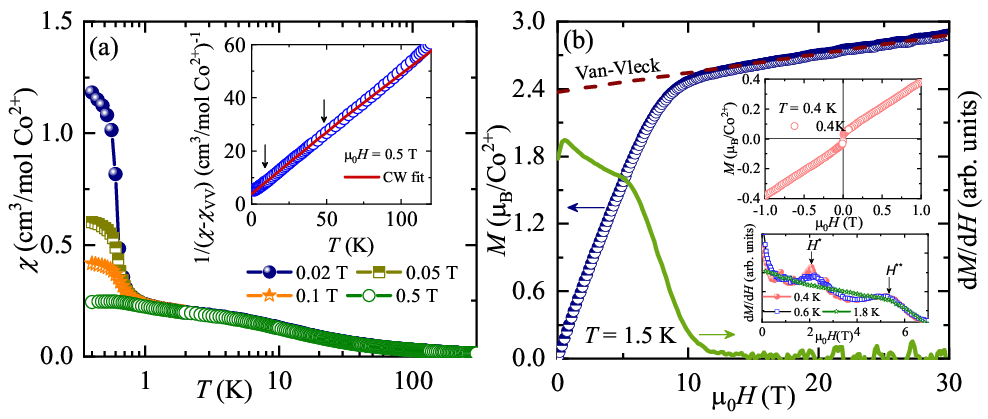}
	\caption{(a) Magnetic susceptibility ($\chi$) vs $T$ in different magnetic fields. Inset: $1/(\chi-\chi_{\rm VV})$ vs $T$ in the low-$T$ region and solid line is the CW fit. The downward arrows indicate the fitting range. $\chi_{\rm VV}$ is the temperature-independent van Vleck contribution determined from the slope of $M(H)$ in high fields. (b) Pulse-field magnetization ($M$) measured at $T = 1.5$~K and its derivative as a function of magnetic field on the left and right axes, respectively. The dashed line is the linear fit to obtain $\chi_{\rm VV}$. Upper inset: $M$ vs $H$ at $T = 0.4$~K. Lower inset: $dM/dH$ vs $H$.}
	\label{Fig2}
\end{figure}
Temperature-dependent magnetic susceptibility $\chi(T)$ measured down to 0.4~K and in different applied magnetic fields on a good quality crystal is presented in Fig.~\ref{Fig2}(a). 
The measurements repeated on the polycrystalline sample almost overlap with the single-crystal data, reflecting no significant anisotropy.
$\chi(T)$ exhibits a sharp increase below $\sim 0.8$~K indicating the onset of a magnetic ordering. The $d\chi/dT$ vs $T$ plot highlights this transition at $T^{*} \sim 0.6$~K (see SM)~\cite{supplementary}. Interestingly, $\chi(T)$ in the ordered state levels off instead of showing a peak, which is typically expected in conventional antiferromagnets. Furthermore, as the magnetic field is increased, the absolute value of $\chi(T)$ below the ordering is suppressed drastically and systematically. All these features point towards either a ferromagnetic or canted antiferromagnetic ordered state.
This transition shifts weakly towards low temperatures before it is suppressed completely above 0.5~T. This gives an indication for a field-induced disordered state appearing in CoKS already in a small applied field. 

To extract the magnetic parameters, the inverse susceptibility ($1/\chi$) for $\mu_0 H = 0.5$~T in high- and low-temperature regions was fitted by the modified Curie-Weiss (CW) law, $\chi(T)=\chi_0+C/(T-\theta_{\rm CW})$. In the high-temperature (HT) region, the Curie constant $C$ corresponds to the effective moment $\mu^{\rm HT}_{\rm eff} \sim 5.8\mu_{\rm B}$, which is higher than the expected spin-only value for $S=3/2$, possibly due to an additional orbital contribution, similar to other Co$^{2+}$ based magnets~\cite{Lal014429,Ranjith115804}.
As evident from Fig.~\ref{Fig2}(a), $\chi(T)$ manifests a clear slope change below about 70~K. This is due to a spin crossover of Co$^{2+}$ from the high-temperature $S = 3/2$ state to an effective pseudospin-$\frac12$ ground state~\cite{Sebastian034403,Lal014429}. Indeed, by fitting the data in the range 5~K~$< T < 50$~K, one finds $C \simeq 2.1$~cm$^3$K/mol-Co$^{2+}$ and the effective moment $\mu^{\rm LT}_{\rm eff} \simeq 4.09$~$\mu_{\rm B}$ that corresponds to $J_{\rm eff}=1/2$ with $g \simeq 4.7$, as typical for Co$^{2+}$ at low temperatures. This $g$-value is independenly confirmed by ESR~\cite{supplementary}. The negative Curie-Weiss temperature, $\theta^{\rm LT}_{\rm CW}$, indicates dominant AFM interactions between the $J_{\rm eff}=1/2$ spins. The resulting frustration ratio of $f = |\theta^{\rm LT}_{\rm CW}|/T^* \sim 13$ pinpoints a strong suppression of the magnetic order in CoKS.


Figure~\ref{Fig2}(b) presents the high-field magnetization ($M$ vs $H$) data recorded at $T \simeq 1.5$~K in pulsed fields up to 30~T. The high-field data are scaled with respect to the SQUID data taken at $T \simeq 1.8$~K up to 7~T. The magnetization increases linearly with field as expected for a typical AFM system and then saturates above $\mu_0H_{\rm S} \sim 10$~T. The saturation field can be pinpointed using the derivative of magnetization ($dM/dH$) vs $H$ plotted in the right $y$-axis where $dM/dH$ value approaches zero. One can also clearly see two distinct features at $H^* \sim 2$~T and $H^{**}\sim 5.3$~T. At $T = 0.4$~K, these features are more pronounced and diminish gradually with increasing temperature [lower inset of Fig.~\ref{Fig2}(b)]. Interestingly, these two anomalies correspond to about $1/3$ and $2/3$ of the saturation magnetization. Weak magnetic couplings in CoKS render these features accessible even in static fields. In the high-field regime ($H > H_{\rm S}$), $M$ shows a linear increase likely due to the Van Vleck contribution of the Co$^{2+}$ ion. The $y$-intercept and slope of the linear fit for $\mu_0H > 18$~T provide the saturation magnetization $M_{\rm S} \sim 2.37 \mu_{\rm B}$ and Van Vleck susceptibility $\chi_{\rm VV} \sim 9.4 \times 10^{-3}$~cm$^3$/mol, respectively. The $M_{\rm S}$ value is close to the expected saturation magnetization for a $J_{\rm eff}=1/2$ system with $g \sim 4.7$. 
The magnetic isotherm measured at $T = 0.4$~K ($<T^*$), does not show any hysteresis [upper inset of Fig.~\ref{Fig2}(b)], although one observes a small net magnetization of 0.04\,$\mu_B$/Co$^{2+}$ that correlates with the susceptibility upturn below $T^*$.
\begin{figure}
\includegraphics[width=\columnwidth]{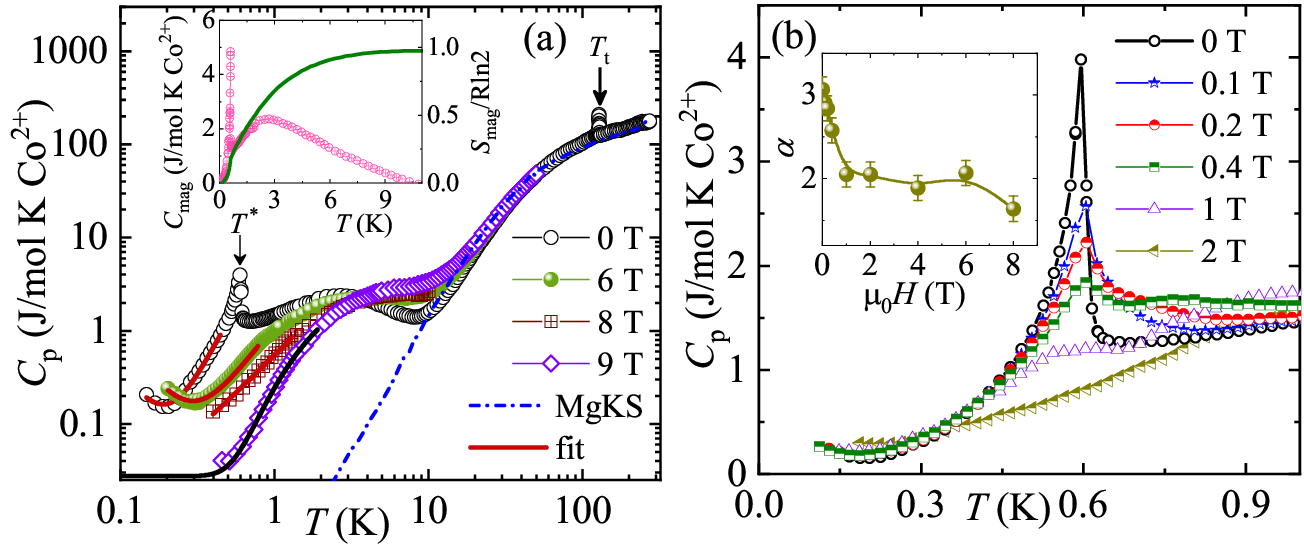}
\caption{(a) Heat capacity ($C_{\rm p}$) vs $T$ measured in zero and higher fields. The dash-dotted line is the phonon contribution from the non-magnetic analog compound K$_2$Mg$_2$(SO$_4$)$_3$. The solid lines are the fits described in the text. For the 9~T data, the solid line represents the exponential fit. Inset: $C_{\rm mag}$ and normalized entropy $S_{\rm mag}/R\ln2$ vs $T$ on left and right axes, respectively. (b) $C_{\rm p}$ vs $T$ measured in low fields highlighting the low-temperature region. Inset: power law exponent $\alpha$ as a function of field.}
	\label{Fig3}
\end{figure}
Temperature-dependent heat capacity [$C_{\rm p}(T)$] measured on a single crystal in different fields down to 100~mK is shown in Fig.~\ref{Fig3}(a). It depicts a sharp anomaly at $T_{\rm t}\simeq 125$~K (see also SM~\cite{supplementary}), reminiscent of a first-order structural transition. The transition temperature determined from thermodynamic measurements is about 11\,K higher than in synchrotron XRD, likely because of the sample heating by the intense x-ray beam in the synchrotron experiment. Below $T_t$, zero-field $C_{\rm p}(T)$ features a broad maximum at around $\sim 10$~K and a $\lambda$-type anomaly at $T^{*} \simeq 0.6$~K.
In order to extract the magnetic part of the heat capacity ($C_{\rm mag}$), we subtracted the phonon contribution by measuring the heat capacity of an isostructural non-magnetic analog compound K$_2$Mg$_2$(SO$_4$)$_3$ (MgKS) that is shown as a dash-dotted line in Fig.~\ref{Fig3}(a). 
The resulting magnetic entropy 
$S_{\rm mag}$ reaches the value of $\sim 5.6$~J/mol-K-Co$^{2+}$ above 10~K, which is very close to $S_{\rm{mag}}=R\ln2 = 5.76$~J/mol-K for $J_{\rm eff}=1/2$. These data provide strong thermodynamic evidence for the formation of the $J_{\rm eff}=1/2$ state of Co$^{2+}$ in CoKS and, hence, for the realization of quantum magnetism in the double-trillium-lattice geometry. 

With the application of a small field, the peak amplitude at $T^{*}\simeq 0.6$~K is considerably reduced and the peak position is shifted very weakly and is completely suppressed above 1~T as shown in the Fig.~\ref{Fig3}(b). This indicates the formation of a field-induced dynamic state, similar to NiKS~\cite{Ivica157204}. In order to access the field-induced behavior, we fitted the $C_{\rm p}$ data at low temperatures ($T < 0.45$~K) by $C_{\rm p}(T)= A/T^2+BT^{\alpha}$. The first term accounts for the nuclear Schottky part as a slight upturn appears at $T < 0.2$~K, while the second term stands for the power-law behavior. The fits are shown in Fig.~\ref{Fig3}(a), and the field evolution of the exponent $\alpha$ is plotted in the inset of Fig.~\ref{Fig3}(b). Clearly, $\alpha$ has a value close to $\sim 3$ in zero field, as expected for a three-dimensional AFM LRO state due to spin-wave dispersion. With an applied field, $\alpha$ decreases rapidly and reaches an almost constant value of $\alpha \sim 2$ for $\mu_0 H > 1$~T. Such a sizable reduction in $\alpha$ has been observed in many frustrated magnets~\cite{Plumb54,Satoru1697}. Unlike the other field values, for $\mu_0 H = 9$~T, $C_p$ falls swiftly towards zero with reduction in temperature. This exponential decay can be ascribed to the gapped magnons in the fully polarized state. 
In Fig.~\ref{Fig3}(a), the fit using $C_{\rm p} = A\,e^{-\Delta/k_{\rm B}T}$ below 2~K yields a magnon gap of $\Delta/k_{\rm B} \simeq 3.2$~K in the fully polarized state at 9\,T.
Furthermore, the position of the broad maximum at $T^{\rm max} \sim 3$~K remains almost unchanged up to 6~T and then shifts slightly towards high temperatures in higher fields [see Fig.~\ref{Fig3}(a)]. This is in contrast to a conventional antiferromagnetic short-range order. A similar feature has been observed in highly frustrated 3D magnets with the QSL ground state~\cite{Ivica157204,Plumb54}. 

\begin{figure}
\includegraphics[width=\columnwidth]{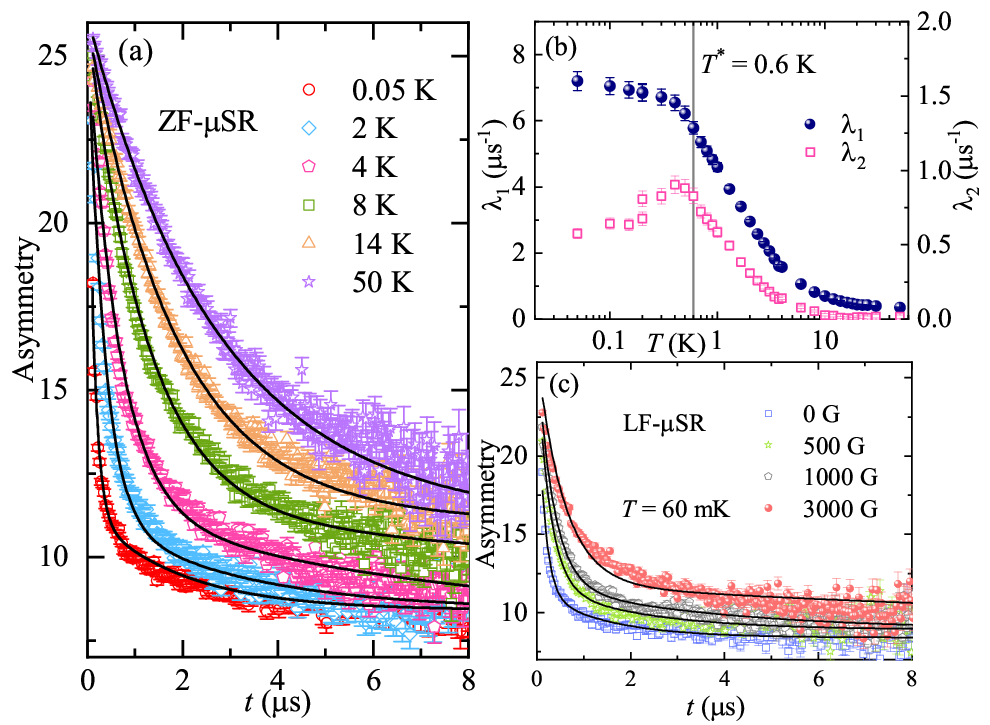}
\caption{(a) ZF-$\mu$SR asymmetries at different temperatures along with the fit by Eq.~\eqref{musr}. (b) The temperature dependence of ZF $\lambda_1$ and $\lambda_2$. (c) $\mu$SR asymmetries at $T = 60$~mK in different LFs with the solid lines as the fit by Eq.~\eqref{musr}.}
\label{Fig4}
\end{figure}
In order to directly examine the spin dynamics, $\mu$SR experiments were performed down to 50~mK. Being highly sensitive to low-energy spin dynamics, $\mu$SR is an excellent local probe of static and dynamic properties of a spin system, especially with fast fluctuations~\cite{Somesh104422}.
Figure~\ref{Fig4}(a) shows the $\mu$SR asymmetries of CoKS at different temperatures measured in zero field. Down to 50~mK, they feature neither any oscillations nor a 1/3$^{\rm rd}$ tail that would be expected in the case of a magnetic LRO or spin freezing, respectively. A slight loss in the initial asymmetry is due to the limited resolution of the MuSR spectrometer at short times and should not be considered as a signature of static order. These observations are quite surprising as the signatures of magnetic LRO could be clearly seen in the magnetization and heat capacity data for CoKS. 

To evaluate the relaxation parameters, we tried to fit the asymmetries by different functions. Since a single exponential function couldn't describe the time-dependent asymmetries satisfactorily, we used a modified function
\begin{equation}
	A(t)= A_0 [fe^{-\lambda_1t}+(1-f)e^{-\lambda_2t}]+A_{\rm bg}.
	\label{musr}
\end{equation} 
Here, $A_0$ is the initial asymmetry, $\lambda_1$ and $\lambda_2$ are the muon relaxation rates, $f$ and ($1-f$) represent their phase fractions, and $A_{\rm bg}$ is the background contribution (constant) from muons stopping at the sample holder. To avoid over-parameterization, we fixed $A_0$ to a value obtained from the high-$T$ asymmetry and $A_{\rm bg} = 8.4$ during the analysis, leaving $f$, $\lambda_1$, and $\lambda_2$ as free parameters.
The two relaxation terms in Eq.~\eqref{musr} were necessary for an accurate description of the data over the entire temperature range. Moreover, we did not observe any Gaussian component in the ZF asymmetry which would normally appear due to static nuclear fields. Hence, Eq.~\eqref{musr} was sufficient to reproduce the experimental data. A similar function was used for fitting the asymmetry of several systems when multiple relaxations were observed due to several muon stopping sites or complex magnetic orders~\cite{Bandyopadhyay184403,Ivica157204}. In the present case, there is a decent possibility of multiple muon sites due to the monoclinic superstructure formed below $T_t$.

The extracted relaxation rates $\lambda_1$ and $\lambda_2$ as a function of temperature are presented in the left and right $y$-axes of Fig.~\ref{Fig4}(b), respectively. Both of them are temperature-independent down to 10~K and show a rapid rise upon further cooling. Such an enhancement signals the slowing down of the spin fluctuations as correlations start to build up, anticipated from the value of $\theta^{\rm LT}_{\rm CW} \simeq -8$~K. Below $T^*\sim 0.6$~K, $\lambda_1$ and $\lambda_2$ show different behaviors. The fast fluctuating component $\lambda_1$ shows a plateau, reflecting persistent spin fluctuations, typically expected in a QSL state~\cite{Li097201,BhattacharyaL060403,Mendels077204}. By contrast, the slow component $\lambda_2$ shows a peak indicating the formation of static order below $T^*$. We found that the phase fraction ($f$) of the fast relaxing component ($\lambda_1$) is about 80~\%. In this scenario, $\lambda_2$ can be associated with a partial ordering of a small fraction of spins, whereas their main fraction still remains in a highly dynamic state. The absence of oscillations can then be viewed as a spread of local fields, possibly due to the complex magnetic structure below $T^*$.

The $\mu$SR asymmetries measured at $T = 60$~mK in different longitudinal fields (LFs) up to 3000~G are shown in Fig.~\ref{Fig4}(c). The asymmetry changes systematically and very slowly with increasing field. It can be fitted well by the same double exponential function [Eq.~\eqref{musr}]. Generally, a LF as high as ten times $\lambda/\gamma_{\mu}$ is sufficient to decouple the muon spin from the local moments, where $\gamma_{\mu}=135.5$~MHz/T is the gyromagnetic ratio of $\mu^{+}$~\cite{Cho014439}. For CoKS, we got $10 \times \lambda_1/\gamma_{\mu}\sim 700$~G, where $\lambda_1 \sim 6$~$\mu$s$^{-1}$ is the ZF relaxation rate at $T = 60$~mK. On the contrary, in the present case, the LF as high as 3000~G (more than four times of 700~G) shows only a partial decoupling. This is another evidence of the presence of strong dynamics (or absence of magnetic order) at a very low temperature of $T = 60$~mK, which is 1/10$^{\rm th}$ of $T^{*}\sim 0.6$~K.
Indeed, similar behavior in LFs is reported for a number of QSL candidates~\cite{Li097201}.


\begin{table}
\caption{\label{tab:exchange}
Exchange couplings $J_i$ (in\,K) calculated for the cubic structure of CoKS. Exchange couplings in NiKS are given for comparison using the notation of Ref.~\cite{Ivica157204}. The interatomic distances are quoted for CoKS (in\,\r A).
}
\begin{ruledtabular}
\begin{tabular}{cccccc}
                        & $J_1$ & $J_2$ & $J_3$ & $J_4$ & $J_5$ \smallskip \\
$d_{\rm Co-Co}$         & 4.460 & 4.935 & 6.126 & 6.163 & 6.170 \\
CoKS                    &  1.0  & $-0.2$ & 1.2  &  1.6  & 2.6\smallskip\\
NiKS~\cite{Ivica157204} &  0.4  & $-0.2$ & 1.1  &  5.4  & 2.5 \\
\end{tabular}
\end{ruledtabular}
\end{table}

Using density-functional (DFT) calculations, we estimated magnetic couplings in CoKS by employing the structure of the cubic phase (at 140\,K, right above $T_t$) for an easier comparison with NiKS~\cite{Ivica157204}. Similar to that compound, we extract five exchange couplings in the double-trillium lattice (Table~\ref{tab:exchange}). All of them are AFM and relatively weak. The CW temperatures are estimated for the Co1 and Co2 sites as
$\theta_{\rm Co1} =-\frac14(J_1+3J_2+6J_3+3J_4)=-3.1\,{\rm K}$ and $\theta_{\rm Co2} =-\frac14(J_1+3J_2+6J_5+3J_4)=-5.2\,{\rm K}$,
resulting in the average value of $-4$\,K in a reasonable agreement with the experimental $\theta^{\rm LT}_{\rm CW} \simeq -8$~K. This coupling regime is quite similar to NiKS, except that $J_5$ becomes the dominant coupling in the Co case, and Co$^{2+}$ is a pseudospin-$\frac12$ ion, in contrast to $S=1$ in Ni$^{2+}$. The AFM nature of $J_1-J_5$ is in sharp contrast to KBaCr$_2$(PO$_4)_3$ where ferromagnetic $J_3$ and $J_5$ eliminate frustration of the double trillium lattice~\cite{Kolay224405}. This difference can be traced back to the absence of the magnetic $e_g$ orbitals in Cr$^{3+}$. By contrast, the $e_g$ orbitals contribute to the magnetic exchange in both Co$^{2+}$ and Ni$^{2+}$, thus leading to the sizable AFM couplings, especially $J_5$.

While the exchange parameters from Table~\ref{tab:exchange} define an effective magnetic model of CoKS and establish its similarity to NiKS, the complete magnetic model of CoKS may be more complex. The superstructure formed below $T_t$ splits each of the couplings $J_1-J_5$ into several nonequivalent terms. Moreover, the couplings $J_{ij}$ between the pseudospin-$\frac12$ Co$^{2+}$ ions may be anisotropic. In view of the complex crystal structure with 228 atoms in the monoclinic unit cell, the derivation of the complete model is quite challenging and should be a subject of a separate dedicated investigation, but already at this stage we can conclude that CoKS features strongly frustrated AFM couplings on the double-trillium lattice and lies close to the regime explored in the previous theoretical studies~\cite{Gonzalez7191}.


Our data indicate proximate QSL behavior in CoKS. Whereas thermodynamic measurements reveal static magnetic order below $T^*\sim 0.6$~K, the absence of oscillations in ZF $\mu$SR asymmetry and the temperature-dependent $\mu$SR relaxation rates provide compelling evidence for the persistence of strong spin fluctuations down to 50~mK, coexisting with a small static component. This was further supported by the $\mu$SR asymmetries measured in LFs at $T=60$~mK, where the field as high as 3000~G was not sufficient to decouple the muon spin completely.

Similar to $\alpha$-RuCl$_3$ and other proximate QSL materials, static magnetic order in CoKS is fragile and can be suppressed in the applied magnetic field. Heat-capacity measurements show a drastic change in the low-temperature power-law behavior from $T^3$ typical for a magnetically ordered state with magnon excitations to $T^2$ that may occur in a Dirac QSL with the linear dispersion of low-energy excitations~\cite{Hermele224413,Zhou197201}. Our study delineates the broad field region between 1 and 10\,T where static magnetic order is absent, and QSL physics can be further probed with spectroscopic methods. It is also worth noting that the magnetization isotherms of CoKS show two anomalies at higher fields that probably indicate distinct regimes within this broad QSL region. Microscopic probes can shed further light on the exact nature of these phases.

CoKS appears to be rather similar to its sister compound NiKS, but we also found a considerable difference between them on the quantitative level. In NiKS, the complete suppression of the static component requires the field of about 4~T, while in CoKS a smaller field of $\sim 1$~T is sufficient. This difference can be traced back to the somewhat weaker magnetic couplings (Table~\ref{tab:exchange}) and the smaller local spin. The $J_{\rm eff}=\frac12$ state of Co$^{2+}$ confirmed in our study renders CoKS the prime candidate for studying quantum magnetism of the double trillium lattice. The ratio of the leading exchange couplings, $J_5/J_4=1.6$, puts CoKS into the most promising parameter regime for stabilizing QSL in the double trillium lattice geometry~\cite{Gonzalez7191}.

In summary, we have grown single crystals of an interesting quantum magnet K$_2$Co$_2$(SO$_4$)$_3$ featuring the double trillium lattice geometry, and demonstrated the proximate QSL behavior of this material. At low temperatures, Co$^{2+}$ develops the pseudospin-$\frac12$ ground state. Static order with a small remnant magnetization is formed below $T^* \simeq 0.6$~K. It coexists with the predominant strongly fluctuating spin component, according to the zero-field and longitudinal-field $\mu$SR data. The static component is fully suppressed around 1\,T where the power-law ($T^{\alpha}$) behavior of the heat capacity changes from $\alpha\sim 3$ to $\alpha\sim 2$. This field-induced state spans the broad field range between 1 and 10\,T and shows no thermodynamic signatures of magnetic LRO. CoKS is a promising playground for studying quantum magnetism in the strongly frustrated double trillium lattice geometry. 

\textit{Note added:} During the completion of our work, the crystal growth and magnetoelectric properties of K$_2$Co$_2$(SO$_4$)$_3$ have been reported by Xu et al.~\cite{Xu2400308} who also concluded on the monoclinic symmetry of the low-temperature structure, but did not detect the three-fold superstructure below $T_t$. Their sample shows the magnetic transition around 0.6\,K, similar to our data, but the low-$T$ spin dynamics has not been probed.

\acknowledgments
For financial support, we would like to acknowledge SERB, India bearing sanction Grant No.~CRG/2022/000997 and DST-FIST with Grant No.~SR/FST/PS-II/2018/54(C). This work was partially carried out using the facilities of UGC-DAE CSR, Indore. The work in Leipzig was funded by funded by the Deutsche Forschungsgemeinschaft (DFG, German Research Foundation) -- TRR 360 -- 492547816 (subproject B3). We also acknowledge the support of HLD-HZDR, member of the European Magnetic Field Laboratory (EMFL), and thank ESRF for providing the beamtime for this experiment.


\begin{thebibliography}{41}%
	\makeatletter
	\providecommand \@ifxundefined [1]{%
		\@ifx{#1\undefined}
	}%
	\providecommand \@ifnum [1]{%
		\ifnum #1\expandafter \@firstoftwo
		\else \expandafter \@secondoftwo
		\fi
	}%
	\providecommand \@ifx [1]{%
		\ifx #1\expandafter \@firstoftwo
		\else \expandafter \@secondoftwo
		\fi
	}%
	\providecommand \natexlab [1]{#1}%
	\providecommand \enquote  [1]{``#1''}%
	\providecommand \bibnamefont  [1]{#1}%
	\providecommand \bibfnamefont [1]{#1}%
	\providecommand \citenamefont [1]{#1}%
	\providecommand \href@noop [0]{\@secondoftwo}%
	\providecommand \href [0]{\begingroup \@sanitize@url \@href}%
	\providecommand \@href[1]{\@@startlink{#1}\@@href}%
	\providecommand \@@href[1]{\endgroup#1\@@endlink}%
	\providecommand \@sanitize@url [0]{\catcode `\\12\catcode `\$12\catcode
		`\&12\catcode `\#12\catcode `\^12\catcode `\_12\catcode `\%12\relax}%
	\providecommand \@@startlink[1]{}%
	\providecommand \@@endlink[0]{}%
	\providecommand \url  [0]{\begingroup\@sanitize@url \@url }%
	\providecommand \@url [1]{\endgroup\@href {#1}{\urlprefix }}%
	\providecommand \urlprefix  [0]{URL }%
	\providecommand \Eprint [0]{\href }%
	\providecommand \doibase [0]{https://doi.org/}%
	\providecommand \selectlanguage [0]{\@gobble}%
	\providecommand \bibinfo  [0]{\@secondoftwo}%
	\providecommand \bibfield  [0]{\@secondoftwo}%
	\providecommand \translation [1]{[#1]}%
	\providecommand \BibitemOpen [0]{}%
	\providecommand \bibitemStop [0]{}%
	\providecommand \bibitemNoStop [0]{.\EOS\space}%
	\providecommand \EOS [0]{\spacefactor3000\relax}%
	\providecommand \BibitemShut  [1]{\csname bibitem#1\endcsname}%
	\let\auto@bib@innerbib\@empty
	\bibitem [{\citenamefont {Broholm}\ \emph {et~al.}(2020)\citenamefont
		{Broholm}, \citenamefont {Cava}, \citenamefont {Kivelson}, \citenamefont
		{Nocera}, \citenamefont {Norman},\ and\ \citenamefont
		{Senthil}}]{Broholm6475}%
	\BibitemOpen
	\bibfield  {author} {\bibinfo {author} {\bibfnamefont {C.}~\bibnamefont
			{Broholm}}, \bibinfo {author} {\bibfnamefont {R.~J.}\ \bibnamefont {Cava}},
		\bibinfo {author} {\bibfnamefont {S.~A.}\ \bibnamefont {Kivelson}}, \bibinfo
		{author} {\bibfnamefont {D.~G.}\ \bibnamefont {Nocera}}, \bibinfo {author}
		{\bibfnamefont {M.~R.}\ \bibnamefont {Norman}},\ and\ \bibinfo {author}
		{\bibfnamefont {T.}~\bibnamefont {Senthil}},\ }\bibfield  {title} {\bibinfo
		{title} {Quantum spin liquids},\ }\href
	{https://doi.org/10.1126/science.aay0668} {\bibfield  {journal} {\bibinfo
			{journal} {Science}\ }\textbf {\bibinfo {volume} {367}},\ \bibinfo {pages}
		{eaay0668} (\bibinfo {year} {2020})}\BibitemShut {NoStop}%
	\bibitem [{\citenamefont {Chamorro}\ \emph {et~al.}(2021)\citenamefont
		{Chamorro}, \citenamefont {McQueen},\ and\ \citenamefont
		{Tran}}]{chamorro2021}%
	\BibitemOpen
	\bibfield  {author} {\bibinfo {author} {\bibfnamefont {J.~R.}\ \bibnamefont
			{Chamorro}}, \bibinfo {author} {\bibfnamefont {T.~M.}\ \bibnamefont
			{McQueen}},\ and\ \bibinfo {author} {\bibfnamefont {T.~T.}\ \bibnamefont
			{Tran}},\ }\bibfield  {title} {\bibinfo {title} {Chemistry of quantum spin
			liquids},\ }\href {https://doi.org/10.1021/acs.chemrev.0c00641} {\bibfield
		{journal} {\bibinfo  {journal} {Chem. Rev.}\ }\textbf {\bibinfo {volume}
			{121}},\ \bibinfo {pages} {2898} (\bibinfo {year} {2021})}\BibitemShut
	{NoStop}%
	\bibitem [{\citenamefont {Chillal}\ \emph {et~al.}(2020)\citenamefont
		{Chillal}, \citenamefont {Iqbal}, \citenamefont {Jeschke}, \citenamefont
		{Rodriguez-Rivera}, \citenamefont {Bewley}, \citenamefont {Manuel},
		\citenamefont {Khalyavin}, \citenamefont {Steffens}, \citenamefont {Thomale},
		\citenamefont {Islam}, \citenamefont {Reuther},\ and\ \citenamefont
		{Lake}}]{Chillal2020}%
	\BibitemOpen
	\bibfield  {author} {\bibinfo {author} {\bibfnamefont {S.}~\bibnamefont
			{Chillal}}, \bibinfo {author} {\bibfnamefont {Y.}~\bibnamefont {Iqbal}},
		\bibinfo {author} {\bibfnamefont {H.~O.}\ \bibnamefont {Jeschke}}, \bibinfo
		{author} {\bibfnamefont {J.~A.}\ \bibnamefont {Rodriguez-Rivera}}, \bibinfo
		{author} {\bibfnamefont {R.}~\bibnamefont {Bewley}}, \bibinfo {author}
		{\bibfnamefont {P.}~\bibnamefont {Manuel}}, \bibinfo {author} {\bibfnamefont
			{D.}~\bibnamefont {Khalyavin}}, \bibinfo {author} {\bibfnamefont
			{P.}~\bibnamefont {Steffens}}, \bibinfo {author} {\bibfnamefont
			{R.}~\bibnamefont {Thomale}}, \bibinfo {author} {\bibfnamefont {A.~T. M.~N.}\
			\bibnamefont {Islam}}, \bibinfo {author} {\bibfnamefont {J.}~\bibnamefont
			{Reuther}},\ and\ \bibinfo {author} {\bibfnamefont {B.}~\bibnamefont
			{Lake}},\ }\bibfield  {title} {\bibinfo {title} {{Evidence for a
				three-dimensional quantum spin liquid in PbCuTe$_2$O$_6$}},\ }\href
	{https://doi.org/10.1038/s41467-020-15594-1} {\bibfield  {journal} {\bibinfo
			{journal} {Nat. Commun.}\ }\textbf {\bibinfo {volume} {11}},\ \bibinfo
		{pages} {2348} (\bibinfo {year} {2020})}\BibitemShut {NoStop}%
	\bibitem [{\citenamefont {Hikita}\ \emph {et~al.}(1977)\citenamefont {Hikita},
		\citenamefont {Sekiguchi},\ and\ \citenamefont {Ikeda}}]{Hikita1327}%
	\BibitemOpen
	\bibfield  {author} {\bibinfo {author} {\bibfnamefont {T.}~\bibnamefont
			{Hikita}}, \bibinfo {author} {\bibfnamefont {H.}~\bibnamefont {Sekiguchi}},\
		and\ \bibinfo {author} {\bibfnamefont {T.}~\bibnamefont {Ikeda}},\ }\bibfield
	{title} {\bibinfo {title} {{Phase Transitions in New Langbeinite-Type
				Crystals}},\ }\href {https://doi.org/10.1143/JPSJ.43.1327} {\bibfield
		{journal} {\bibinfo  {journal} {J. Phys. Soc. Jpn.}\ }\textbf {\bibinfo
			{volume} {43}},\ \bibinfo {pages} {1327} (\bibinfo {year}
		{1977})}\BibitemShut {NoStop}%
	\bibitem [{\citenamefont {Percival}\ and\ \citenamefont
		{Salje}(1989)}]{Percival563}%
	\BibitemOpen
	\bibfield  {author} {\bibinfo {author} {\bibfnamefont {M.}~\bibnamefont
			{Percival}}\ and\ \bibinfo {author} {\bibfnamefont {E.}~\bibnamefont
			{Salje}},\ }\bibfield  {title} {\bibinfo {title} {{Optical absorption
				spectroscopy of the $P2_13-P2_12_12_1$ transformation in
				K$_2$Co$_2$(SO$_4$)$_3$ langbeinite}},\ }\href@noop {} {\bibfield  {journal}
		{\bibinfo  {journal} {Phys. Chem. Miner.}\ }\textbf {\bibinfo {volume}
			{16}},\ \bibinfo {pages} {563} (\bibinfo {year} {1989})}\BibitemShut
	{NoStop}%
	\bibitem [{\citenamefont {\ifmmode \check{Z}\else
			\v{Z}\fi{}ivkovi\ifmmode~\acute{c}\else \'{c}\fi{}}\ \emph
		{et~al.}(2021)\citenamefont {\ifmmode \check{Z}\else
			\v{Z}\fi{}ivkovi\ifmmode~\acute{c}\else \'{c}\fi{}}, \citenamefont {Favre},
		\citenamefont {Salazar~Mejia}, \citenamefont {Jeschke}, \citenamefont
		{Magrez}, \citenamefont {Dabholkar}, \citenamefont {Noculak}, \citenamefont
		{Freitas}, \citenamefont {Jeong}, \citenamefont {Hegde}, \citenamefont
		{Testa}, \citenamefont {Babkevich}, \citenamefont {Su}, \citenamefont
		{Manuel}, \citenamefont {Luetkens}, \citenamefont {Baines}, \citenamefont
		{Baker}, \citenamefont {Wosnitza}, \citenamefont {Zaharko}, \citenamefont
		{Iqbal}, \citenamefont {Reuther},\ and\ \citenamefont
		{R\o{}nnow}}]{Ivica157204}%
	\BibitemOpen
	\bibfield  {author} {\bibinfo {author} {\bibfnamefont {I.}~\bibnamefont
			{\ifmmode \check{Z}\else \v{Z}\fi{}ivkovi\ifmmode~\acute{c}\else
				\'{c}\fi{}}}, \bibinfo {author} {\bibfnamefont {V.}~\bibnamefont {Favre}},
		\bibinfo {author} {\bibfnamefont {C.}~\bibnamefont {Salazar~Mejia}}, \bibinfo
		{author} {\bibfnamefont {H.~O.}\ \bibnamefont {Jeschke}}, \bibinfo {author}
		{\bibfnamefont {A.}~\bibnamefont {Magrez}}, \bibinfo {author} {\bibfnamefont
			{B.}~\bibnamefont {Dabholkar}}, \bibinfo {author} {\bibfnamefont
			{V.}~\bibnamefont {Noculak}}, \bibinfo {author} {\bibfnamefont {R.~S.}\
			\bibnamefont {Freitas}}, \bibinfo {author} {\bibfnamefont {M.}~\bibnamefont
			{Jeong}}, \bibinfo {author} {\bibfnamefont {N.~G.}\ \bibnamefont {Hegde}},
		\bibinfo {author} {\bibfnamefont {L.}~\bibnamefont {Testa}}, \bibinfo
		{author} {\bibfnamefont {P.}~\bibnamefont {Babkevich}}, \bibinfo {author}
		{\bibfnamefont {Y.}~\bibnamefont {Su}}, \bibinfo {author} {\bibfnamefont
			{P.}~\bibnamefont {Manuel}}, \bibinfo {author} {\bibfnamefont
			{H.}~\bibnamefont {Luetkens}}, \bibinfo {author} {\bibfnamefont
			{C.}~\bibnamefont {Baines}}, \bibinfo {author} {\bibfnamefont {P.~J.}\
			\bibnamefont {Baker}}, \bibinfo {author} {\bibfnamefont {J.}~\bibnamefont
			{Wosnitza}}, \bibinfo {author} {\bibfnamefont {O.}~\bibnamefont {Zaharko}},
		\bibinfo {author} {\bibfnamefont {Y.}~\bibnamefont {Iqbal}}, \bibinfo
		{author} {\bibfnamefont {J.}~\bibnamefont {Reuther}},\ and\ \bibinfo {author}
		{\bibfnamefont {H.~M.}\ \bibnamefont {R\o{}nnow}},\ }\bibfield  {title}
	{\bibinfo {title} {{Magnetic Field Induced Quantum Spin Liquid in the Two
				Coupled Trillium Lattices of
				${\mathrm{K}}_{2}{\mathrm{Ni}}_{2}({\mathrm{SO}}_{4}{)}_{3}$}},\ }\href
	{https://doi.org/10.1103/PhysRevLett.127.157204} {\bibfield  {journal}
		{\bibinfo  {journal} {Phys. Rev. Lett.}\ }\textbf {\bibinfo {volume} {127}},\
		\bibinfo {pages} {157204} (\bibinfo {year} {2021})}\BibitemShut {NoStop}%
	\bibitem [{\citenamefont {Yao}\ \emph {et~al.}(2023)\citenamefont {Yao},
		\citenamefont {Huang}, \citenamefont {Xie}, \citenamefont {Podlesnyak},
		\citenamefont {Brassington}, \citenamefont {Xing}, \citenamefont
		{Mudiyanselage}, \citenamefont {Wang}, \citenamefont {Xie}, \citenamefont
		{Zhang}, \citenamefont {Lee}, \citenamefont {Zapf}, \citenamefont {Bai},
		\citenamefont {Tennant}, \citenamefont {Liu},\ and\ \citenamefont
		{Zhou}}]{Yao146701}%
	\BibitemOpen
	\bibfield  {author} {\bibinfo {author} {\bibfnamefont {W.}~\bibnamefont
			{Yao}}, \bibinfo {author} {\bibfnamefont {Q.}~\bibnamefont {Huang}}, \bibinfo
		{author} {\bibfnamefont {T.}~\bibnamefont {Xie}}, \bibinfo {author}
		{\bibfnamefont {A.}~\bibnamefont {Podlesnyak}}, \bibinfo {author}
		{\bibfnamefont {A.}~\bibnamefont {Brassington}}, \bibinfo {author}
		{\bibfnamefont {C.}~\bibnamefont {Xing}}, \bibinfo {author} {\bibfnamefont
			{R.~S.~D.}\ \bibnamefont {Mudiyanselage}}, \bibinfo {author} {\bibfnamefont
			{H.}~\bibnamefont {Wang}}, \bibinfo {author} {\bibfnamefont {W.}~\bibnamefont
			{Xie}}, \bibinfo {author} {\bibfnamefont {S.}~\bibnamefont {Zhang}}, \bibinfo
		{author} {\bibfnamefont {M.}~\bibnamefont {Lee}}, \bibinfo {author}
		{\bibfnamefont {V.~S.}\ \bibnamefont {Zapf}}, \bibinfo {author}
		{\bibfnamefont {X.}~\bibnamefont {Bai}}, \bibinfo {author} {\bibfnamefont
			{D.~A.}\ \bibnamefont {Tennant}}, \bibinfo {author} {\bibfnamefont
			{J.}~\bibnamefont {Liu}},\ and\ \bibinfo {author} {\bibfnamefont
			{H.}~\bibnamefont {Zhou}},\ }\bibfield  {title} {\bibinfo {title}
		{{Continuous Spin Excitations in the Three-Dimensional Frustrated Magnet
				${\mathrm{K}}_{2}{\mathrm{Ni}}_{2}({\mathrm{SO}}_{4}{)}_{3}$}},\ }\href
	{https://doi.org/10.1103/PhysRevLett.131.146701} {\bibfield  {journal}
		{\bibinfo  {journal} {Phys. Rev. Lett.}\ }\textbf {\bibinfo {volume} {131}},\
		\bibinfo {pages} {146701} (\bibinfo {year} {2023})}\BibitemShut {NoStop}%
	\bibitem [{\citenamefont {Gonzalez}\ \emph {et~al.}(2024)\citenamefont
		{Gonzalez}, \citenamefont {Noculak}, \citenamefont {Sharma}, \citenamefont
		{Favre}, \citenamefont {Soh}, \citenamefont {Magrez}, \citenamefont {Bewley},
		\citenamefont {Jeschke}, \citenamefont {Reuther}, \citenamefont {R{\o}nnow},
		\citenamefont {Iqbal},\ and\ \citenamefont
		{{\v{Z}}ivkovi{\'{c}}}}]{Gonzalez7191}%
	\BibitemOpen
	\bibfield  {author} {\bibinfo {author} {\bibfnamefont {M.~G.}\ \bibnamefont
			{Gonzalez}}, \bibinfo {author} {\bibfnamefont {V.}~\bibnamefont {Noculak}},
		\bibinfo {author} {\bibfnamefont {A.}~\bibnamefont {Sharma}}, \bibinfo
		{author} {\bibfnamefont {V.}~\bibnamefont {Favre}}, \bibinfo {author}
		{\bibfnamefont {J.-R.}\ \bibnamefont {Soh}}, \bibinfo {author} {\bibfnamefont
			{A.}~\bibnamefont {Magrez}}, \bibinfo {author} {\bibfnamefont
			{R.}~\bibnamefont {Bewley}}, \bibinfo {author} {\bibfnamefont {H.~O.}\
			\bibnamefont {Jeschke}}, \bibinfo {author} {\bibfnamefont {J.}~\bibnamefont
			{Reuther}}, \bibinfo {author} {\bibfnamefont {H.~M.}\ \bibnamefont
			{R{\o}nnow}}, \bibinfo {author} {\bibfnamefont {Y.}~\bibnamefont {Iqbal}},\
		and\ \bibinfo {author} {\bibfnamefont {I.}~\bibnamefont
			{{\v{Z}}ivkovi{\'{c}}}},\ }\bibfield  {title} {\bibinfo {title} {{Dynamics of
				K$_2$Ni$_2$(SO$_4$)$_3$ governed by proximity to a 3D spin liquid model}},\
	}\href {https://doi.org/10.1038/s41467-024-51362-1} {\bibfield  {journal}
		{\bibinfo  {journal} {Nat. Commun.}\ }\textbf {\bibinfo {volume} {15}},\
		\bibinfo {pages} {7191} (\bibinfo {year} {2024})}\BibitemShut {NoStop}%
	\bibitem [{\citenamefont {Boya}\ \emph {et~al.}(2022)\citenamefont {Boya},
		\citenamefont {Nam}, \citenamefont {Kargeti}, \citenamefont {Jain},
		\citenamefont {Kumar}, \citenamefont {Panda}, \citenamefont {Yusuf},
		\citenamefont {Paulose}, \citenamefont {Voma}, \citenamefont {Kermarrec},
		\citenamefont {Kim},\ and\ \citenamefont {Koteswararao}}]{Boya101103}%
	\BibitemOpen
	\bibfield  {author} {\bibinfo {author} {\bibfnamefont {K.}~\bibnamefont
			{Boya}}, \bibinfo {author} {\bibfnamefont {K.}~\bibnamefont {Nam}}, \bibinfo
		{author} {\bibfnamefont {K.}~\bibnamefont {Kargeti}}, \bibinfo {author}
		{\bibfnamefont {A.}~\bibnamefont {Jain}}, \bibinfo {author} {\bibfnamefont
			{R.}~\bibnamefont {Kumar}}, \bibinfo {author} {\bibfnamefont {S.~K.}\
			\bibnamefont {Panda}}, \bibinfo {author} {\bibfnamefont {S.~M.}\ \bibnamefont
			{Yusuf}}, \bibinfo {author} {\bibfnamefont {P.~L.}\ \bibnamefont {Paulose}},
		\bibinfo {author} {\bibfnamefont {U.~K.}\ \bibnamefont {Voma}}, \bibinfo
		{author} {\bibfnamefont {E.}~\bibnamefont {Kermarrec}}, \bibinfo {author}
		{\bibfnamefont {K.~H.}\ \bibnamefont {Kim}},\ and\ \bibinfo {author}
		{\bibfnamefont {B.}~\bibnamefont {Koteswararao}},\ }\bibfield  {title}
	{\bibinfo {title} {{Signatures of spin-liquid state in a 3D frustrated
				lattice compound KSrFe$_2$(PO$_4$)$_3$ with $S = 5/2$}},\ }\href
	{https://doi.org/10.1063/5.0096942} {\bibfield  {journal} {\bibinfo
			{journal} {APL Mater.}\ }\textbf {\bibinfo {volume} {10}},\ \bibinfo {pages}
		{101103} (\bibinfo {year} {2022})}\BibitemShut {NoStop}%
	\bibitem [{\citenamefont {Khatua}\ \emph {et~al.}(2024)\citenamefont {Khatua},
		\citenamefont {Lee}, \citenamefont {Ban}, \citenamefont {Uhlarz},
		\citenamefont {Murugan}, \citenamefont {Sankar}, \citenamefont {Choi},\ and\
		\citenamefont {Khuntia}}]{Khatua184432}%
	\BibitemOpen
	\bibfield  {author} {\bibinfo {author} {\bibfnamefont {J.}~\bibnamefont
			{Khatua}}, \bibinfo {author} {\bibfnamefont {S.}~\bibnamefont {Lee}},
		\bibinfo {author} {\bibfnamefont {G.}~\bibnamefont {Ban}}, \bibinfo {author}
		{\bibfnamefont {M.}~\bibnamefont {Uhlarz}}, \bibinfo {author} {\bibfnamefont
			{G.~S.}\ \bibnamefont {Murugan}}, \bibinfo {author} {\bibfnamefont
			{R.}~\bibnamefont {Sankar}}, \bibinfo {author} {\bibfnamefont {K.-Y.}\
			\bibnamefont {Choi}},\ and\ \bibinfo {author} {\bibfnamefont
			{P.}~\bibnamefont {Khuntia}},\ }\bibfield  {title} {\bibinfo {title}
		{{Magnetism and spin dynamics of the $S$=$\frac{3}{2}$ frustrated trillium
				lattice compound ${\mathrm{K}}_{2}\mathrm{CrTi}({\mathrm{PO}}_{4}{)}_{3}$}},\
	}\href {https://doi.org/10.1103/PhysRevB.109.184432} {\bibfield  {journal}
		{\bibinfo  {journal} {Phys. Rev. B}\ }\textbf {\bibinfo {volume} {109}},\
		\bibinfo {pages} {184432} (\bibinfo {year} {2024})}\BibitemShut {NoStop}%
	\bibitem [{\citenamefont {Kolay}\ \emph {et~al.}(2024)\citenamefont {Kolay},
		\citenamefont {Ding}, \citenamefont {Furukawa}, \citenamefont {Tsirlin},\
		and\ \citenamefont {Nath}}]{Kolay224405}%
	\BibitemOpen
	\bibfield  {author} {\bibinfo {author} {\bibfnamefont {R.}~\bibnamefont
			{Kolay}}, \bibinfo {author} {\bibfnamefont {Q.-P.}\ \bibnamefont {Ding}},
		\bibinfo {author} {\bibfnamefont {Y.}~\bibnamefont {Furukawa}}, \bibinfo
		{author} {\bibfnamefont {A.~A.}\ \bibnamefont {Tsirlin}},\ and\ \bibinfo
		{author} {\bibfnamefont {R.}~\bibnamefont {Nath}},\ }\bibfield  {title}
	{\bibinfo {title} {{Magnetic properties of the double trillium lattice
				antiferromagnet ${\mathrm{KBaCr}}_{2}{({\mathrm{PO}}_{4})}_{3}$}},\ }\href
	{https://doi.org/10.1103/PhysRevB.110.224405} {\bibfield  {journal} {\bibinfo
			{journal} {Phys. Rev. B}\ }\textbf {\bibinfo {volume} {110}},\ \bibinfo
		{pages} {224405} (\bibinfo {year} {2024})}\BibitemShut {NoStop}%
	\bibitem [{\citenamefont {Březina}\ and\ \citenamefont
		{Fousková}(1978)}]{Brezina623}%
	\BibitemOpen
	\bibfield  {author} {\bibinfo {author} {\bibfnamefont {B.}~\bibnamefont
			{Březina}}\ and\ \bibinfo {author} {\bibfnamefont {A.}~\bibnamefont
			{Fousková}},\ }\bibfield  {title} {\bibinfo {title} {{The growth of single
				crystals of langbeinites Rb$_2$Cd$_2$(SO$_4$)$_3$, Tl$_2$Cd$_2$(SO$_4$)$_3$,
				K$_2$Co$_2$(SO$_4$)$_3$, and their phase transitions}},\ }\href
	{https://doi.org/https://doi.org/10.1002/crat.19780130604} {\bibfield
		{journal} {\bibinfo  {journal} {Kristall und Technik}\ }\textbf {\bibinfo
			{volume} {13}},\ \bibinfo {pages} {623} (\bibinfo {year} {1978})}\BibitemShut
	{NoStop}%
	\bibitem [{sup()}]{supplementary}%
	\BibitemOpen
	\href@noop {} {}\bibinfo {note} {See Supplemental Material at http: LINK for
		additional information about methods adopted, single crystal data, powder
		XRD, magnetization, ESR, heat capacity, and thermal conductivity which also
		includes
		Refs.~\cite{Sheldrick2018shelxl,esrf,Fitch1003,Tsirlin132407,Pratt710,vasp1,vasp2,pbe96,tsirlin2014,bader2022,madruga2024,Xu267202}}\BibitemShut
	{NoStop}%
	\bibitem [{\citenamefont {Moriyoshi}\ and\ \citenamefont
		{Hikita}(1995)}]{Moriyoshi4726}%
	\BibitemOpen
	\bibfield  {author} {\bibinfo {author} {\bibfnamefont {K.}~\bibnamefont
			{Moriyoshi}, \bibfnamefont {and~Itoh}}\ and\ \bibinfo {author} {\bibfnamefont
			{T.}~\bibnamefont {Hikita}},\ }\bibfield  {title} {\bibinfo {title}
		{{Structural Study of Phase Transition in K$_2$Co$_2$(SO$_4$)$_3$
				Crystals}},\ }\href {https://doi.org/10.1143/jpsj.64.4726} {\bibfield
		{journal} {\bibinfo  {journal} {J. Phys. Soc. Jpn.}\ }\textbf {\bibinfo
			{volume} {64}},\ \bibinfo {pages} {4726} (\bibinfo {year}
		{1995})}\BibitemShut {NoStop}%
	\bibitem [{\citenamefont {Brown}\ and\ \citenamefont
		{Altermatt}(1985)}]{brown1985}%
	\BibitemOpen
	\bibfield  {author} {\bibinfo {author} {\bibfnamefont {I.~D.}\ \bibnamefont
			{Brown}}\ and\ \bibinfo {author} {\bibfnamefont {D.}~\bibnamefont
			{Altermatt}},\ }\bibfield  {title} {\bibinfo {title} {Bond-valence parameters
			obtained from a systematic analysis of the {Inorganic Crystal Structure
				Database}},\ }\href {https://doi.org/10.1107/S0108768185002063} {\bibfield
		{journal} {\bibinfo  {journal} {Acta Cryst.}\ }\textbf {\bibinfo {volume}
			{B41}},\ \bibinfo {pages} {244} (\bibinfo {year} {1985})}\BibitemShut
	{NoStop}%
	\bibitem [{\citenamefont {Lal}\ \emph {et~al.}(2023)\citenamefont {Lal},
		\citenamefont {Sebastian}, \citenamefont {Islam}, \citenamefont {Saravanan},
		\citenamefont {Uhlarz}, \citenamefont {Skourski},\ and\ \citenamefont
		{Nath}}]{Lal014429}%
	\BibitemOpen
	\bibfield  {author} {\bibinfo {author} {\bibfnamefont {S.}~\bibnamefont
			{Lal}}, \bibinfo {author} {\bibfnamefont {S.~J.}\ \bibnamefont {Sebastian}},
		\bibinfo {author} {\bibfnamefont {S.~S.}\ \bibnamefont {Islam}}, \bibinfo
		{author} {\bibfnamefont {M.~P.}\ \bibnamefont {Saravanan}}, \bibinfo {author}
		{\bibfnamefont {M.}~\bibnamefont {Uhlarz}}, \bibinfo {author} {\bibfnamefont
			{Y.}~\bibnamefont {Skourski}},\ and\ \bibinfo {author} {\bibfnamefont
			{R.}~\bibnamefont {Nath}},\ }\bibfield  {title} {\bibinfo {title} {{Double
				magnetic transitions and exotic field-induced phase in the triangular lattice
				antiferromagnets
				${\mathrm{Sr}}_{3}\mathrm{Co}{(\mathrm{Nb},\mathrm{Ta})}_{2}{\mathrm{O}}_{9}$}},\
	}\href {https://doi.org/10.1103/PhysRevB.108.014429} {\bibfield  {journal}
		{\bibinfo  {journal} {Phys. Rev. B}\ }\textbf {\bibinfo {volume} {108}},\
		\bibinfo {pages} {014429} (\bibinfo {year} {2023})}\BibitemShut {NoStop}%
	\bibitem [{\citenamefont {Ranjith}\ \emph {et~al.}(2017)\citenamefont
		{Ranjith}, \citenamefont {Brinda}, \citenamefont {Arjun}, \citenamefont
		{Hegde},\ and\ \citenamefont {Nath}}]{Ranjith115804}%
	\BibitemOpen
	\bibfield  {author} {\bibinfo {author} {\bibfnamefont {K.~M.}\ \bibnamefont
			{Ranjith}}, \bibinfo {author} {\bibfnamefont {K.}~\bibnamefont {Brinda}},
		\bibinfo {author} {\bibfnamefont {U.}~\bibnamefont {Arjun}}, \bibinfo
		{author} {\bibfnamefont {N.~G.}\ \bibnamefont {Hegde}},\ and\ \bibinfo
		{author} {\bibfnamefont {R.}~\bibnamefont {Nath}},\ }\bibfield  {title}
	{\bibinfo {title} {{Double phase transition in the triangular antiferromagnet
				Ba$_3$CoTa$_2$O$_9$}},\ }\href {https://doi.org/10.1088/1361-648X/aa57be}
	{\bibfield  {journal} {\bibinfo  {journal} {J. Phys.: Condens. Matter}\
		}\textbf {\bibinfo {volume} {29}},\ \bibinfo {pages} {115804} (\bibinfo
		{year} {2017})}\BibitemShut {NoStop}%
	\bibitem [{\citenamefont {Sebastian}\ \emph {et~al.}(2024)\citenamefont
		{Sebastian}, \citenamefont {Mohanty}, \citenamefont {Nath}, \citenamefont
		{Saravanan}, \citenamefont {Mandal}, \citenamefont {Tsirlin},\ and\
		\citenamefont {Nath}}]{Sebastian034403}%
	\BibitemOpen
	\bibfield  {author} {\bibinfo {author} {\bibfnamefont {S.~J.}\ \bibnamefont
			{Sebastian}}, \bibinfo {author} {\bibfnamefont {S.}~\bibnamefont {Mohanty}},
		\bibinfo {author} {\bibfnamefont {A.}~\bibnamefont {Nath}}, \bibinfo {author}
		{\bibfnamefont {M.~P.}\ \bibnamefont {Saravanan}}, \bibinfo {author}
		{\bibfnamefont {S.}~\bibnamefont {Mandal}}, \bibinfo {author} {\bibfnamefont
			{A.~A.}\ \bibnamefont {Tsirlin}},\ and\ \bibinfo {author} {\bibfnamefont
			{R.}~\bibnamefont {Nath}},\ }\bibfield  {title} {\bibinfo {title}
		{{Disordered ground state in a spin-orbit coupled pseudospin-$\frac{1}{2}$
				cobalt-based metal-organic framework magnet with orthogonal spin dimers}},\
	}\href {https://doi.org/10.1103/PhysRevMaterials.8.034403} {\bibfield
		{journal} {\bibinfo  {journal} {Phys. Rev. Mater.}\ }\textbf {\bibinfo
			{volume} {8}},\ \bibinfo {pages} {034403} (\bibinfo {year}
		{2024})}\BibitemShut {NoStop}%
	\bibitem [{\citenamefont {Plumb}\ \emph {et~al.}(2019)\citenamefont {Plumb},
		\citenamefont {Changlani}, \citenamefont {Scheie}, \citenamefont {Zhang},
		\citenamefont {Krizan}, \citenamefont {Rodriguez-Rivera}, \citenamefont
		{Qiu}, \citenamefont {Winn}, \citenamefont {Cava},\ and\ \citenamefont
		{Broholm}}]{Plumb54}%
	\BibitemOpen
	\bibfield  {author} {\bibinfo {author} {\bibfnamefont {K.~W.}\ \bibnamefont
			{Plumb}}, \bibinfo {author} {\bibfnamefont {H.~J.}\ \bibnamefont
			{Changlani}}, \bibinfo {author} {\bibfnamefont {A.}~\bibnamefont {Scheie}},
		\bibinfo {author} {\bibfnamefont {S.}~\bibnamefont {Zhang}}, \bibinfo
		{author} {\bibfnamefont {J.~W.}\ \bibnamefont {Krizan}}, \bibinfo {author}
		{\bibfnamefont {J.~A.}\ \bibnamefont {Rodriguez-Rivera}}, \bibinfo {author}
		{\bibfnamefont {Y.}~\bibnamefont {Qiu}}, \bibinfo {author} {\bibfnamefont
			{B.}~\bibnamefont {Winn}}, \bibinfo {author} {\bibfnamefont {R.~J.}\
			\bibnamefont {Cava}},\ and\ \bibinfo {author} {\bibfnamefont {C.~L.}\
			\bibnamefont {Broholm}},\ }\bibfield  {title} {\bibinfo {title} {{Continuum
				of quantum fluctuations in a three-dimensional $S=1$ Heisenberg magnet}},\
	}\href {https://doi.org/10.1038/s41567-018-0317-3} {\bibfield  {journal}
		{\bibinfo  {journal} {Nat. Phys.}\ }\textbf {\bibinfo {volume} {15}},\
		\bibinfo {pages} {54} (\bibinfo {year} {2019})}\BibitemShut {NoStop}%
	\bibitem [{\citenamefont {Nakatsuji}\ \emph {et~al.}(2005)\citenamefont
		{Nakatsuji}, \citenamefont {Nambu}, \citenamefont {Tonomura}, \citenamefont
		{Sakai}, \citenamefont {Jonas}, \citenamefont {Broholm}, \citenamefont
		{Tsunetsugu}, \citenamefont {Qiu},\ and\ \citenamefont {Maeno}}]{Satoru1697}%
	\BibitemOpen
	\bibfield  {author} {\bibinfo {author} {\bibfnamefont {S.}~\bibnamefont
			{Nakatsuji}}, \bibinfo {author} {\bibfnamefont {Y.}~\bibnamefont {Nambu}},
		\bibinfo {author} {\bibfnamefont {H.}~\bibnamefont {Tonomura}}, \bibinfo
		{author} {\bibfnamefont {O.}~\bibnamefont {Sakai}}, \bibinfo {author}
		{\bibfnamefont {S.}~\bibnamefont {Jonas}}, \bibinfo {author} {\bibfnamefont
			{C.}~\bibnamefont {Broholm}}, \bibinfo {author} {\bibfnamefont
			{H.}~\bibnamefont {Tsunetsugu}}, \bibinfo {author} {\bibfnamefont
			{Y.}~\bibnamefont {Qiu}},\ and\ \bibinfo {author} {\bibfnamefont
			{Y.}~\bibnamefont {Maeno}},\ }\bibfield  {title} {\bibinfo {title} {{Spin
				Disorder on a Triangular Lattice}},\ }\href
	{https://doi.org/10.1126/science.1114727} {\bibfield  {journal} {\bibinfo
			{journal} {Science}\ }\textbf {\bibinfo {volume} {309}},\ \bibinfo {pages}
		{1697} (\bibinfo {year} {2005})}\BibitemShut {NoStop}%
	\bibitem [{\citenamefont {Somesh}\ \emph {et~al.}(2021)\citenamefont {Somesh},
		\citenamefont {Furukawa}, \citenamefont {Simutis}, \citenamefont {Bert},
		\citenamefont {Prinz-Zwick}, \citenamefont {B\"uttgen}, \citenamefont
		{Zorko}, \citenamefont {Tsirlin}, \citenamefont {Mendels},\ and\
		\citenamefont {Nath}}]{Somesh104422}%
	\BibitemOpen
	\bibfield  {author} {\bibinfo {author} {\bibfnamefont {K.}~\bibnamefont
			{Somesh}}, \bibinfo {author} {\bibfnamefont {Y.}~\bibnamefont {Furukawa}},
		\bibinfo {author} {\bibfnamefont {G.}~\bibnamefont {Simutis}}, \bibinfo
		{author} {\bibfnamefont {F.}~\bibnamefont {Bert}}, \bibinfo {author}
		{\bibfnamefont {M.}~\bibnamefont {Prinz-Zwick}}, \bibinfo {author}
		{\bibfnamefont {N.}~\bibnamefont {B\"uttgen}}, \bibinfo {author}
		{\bibfnamefont {A.}~\bibnamefont {Zorko}}, \bibinfo {author} {\bibfnamefont
			{A.~A.}\ \bibnamefont {Tsirlin}}, \bibinfo {author} {\bibfnamefont
			{P.}~\bibnamefont {Mendels}},\ and\ \bibinfo {author} {\bibfnamefont
			{R.}~\bibnamefont {Nath}},\ }\bibfield  {title} {\bibinfo {title} {Universal
			fluctuating regime in triangular chromate antiferromagnets},\ }\href
	{https://doi.org/10.1103/PhysRevB.104.104422} {\bibfield  {journal} {\bibinfo
			{journal} {Phys. Rev. B}\ }\textbf {\bibinfo {volume} {104}},\ \bibinfo
		{pages} {104422} (\bibinfo {year} {2021})}\BibitemShut {NoStop}%
	\bibitem [{\citenamefont {Bandyopadhyay}\ \emph {et~al.}(2024)\citenamefont
		{Bandyopadhyay}, \citenamefont {Lee}, \citenamefont {Adroja}, \citenamefont
		{Stenning}, \citenamefont {Berlie}, \citenamefont {Lees}, \citenamefont
		{Saha}, \citenamefont {Takegami}, \citenamefont {Mel\'endez-Sans},
		\citenamefont {Poelchen}, \citenamefont {Yoshimura}, \citenamefont {Tsuei},
		\citenamefont {Hu}, \citenamefont {Kao}, \citenamefont {Huang}, \citenamefont
		{Chan},\ and\ \citenamefont {Choi}}]{Bandyopadhyay184403}%
	\BibitemOpen
	\bibfield  {author} {\bibinfo {author} {\bibfnamefont {A.}~\bibnamefont
			{Bandyopadhyay}}, \bibinfo {author} {\bibfnamefont {S.}~\bibnamefont {Lee}},
		\bibinfo {author} {\bibfnamefont {D.~T.}\ \bibnamefont {Adroja}}, \bibinfo
		{author} {\bibfnamefont {G.~B.~G.}\ \bibnamefont {Stenning}}, \bibinfo
		{author} {\bibfnamefont {A.}~\bibnamefont {Berlie}}, \bibinfo {author}
		{\bibfnamefont {M.~R.}\ \bibnamefont {Lees}}, \bibinfo {author}
		{\bibfnamefont {R.~A.}\ \bibnamefont {Saha}}, \bibinfo {author}
		{\bibfnamefont {D.}~\bibnamefont {Takegami}}, \bibinfo {author}
		{\bibfnamefont {A.}~\bibnamefont {Mel\'endez-Sans}}, \bibinfo {author}
		{\bibfnamefont {G.}~\bibnamefont {Poelchen}}, \bibinfo {author}
		{\bibfnamefont {M.}~\bibnamefont {Yoshimura}}, \bibinfo {author}
		{\bibfnamefont {K.~D.}\ \bibnamefont {Tsuei}}, \bibinfo {author}
		{\bibfnamefont {Z.}~\bibnamefont {Hu}}, \bibinfo {author} {\bibfnamefont
			{C.-W.}\ \bibnamefont {Kao}}, \bibinfo {author} {\bibfnamefont {Y.-C.}\
			\bibnamefont {Huang}}, \bibinfo {author} {\bibfnamefont {T.-S.}\ \bibnamefont
			{Chan}},\ and\ \bibinfo {author} {\bibfnamefont {K.-Y.}\ \bibnamefont
			{Choi}},\ }\bibfield  {title} {\bibinfo {title} {{Quantum spin liquid ground
				state in the trimer rhodate
				${\mathrm{Ba}}_{4}{\mathrm{NbRh}}_{3}{\mathrm{O}}_{12}$}},\ }\href
	{https://doi.org/10.1103/PhysRevB.109.184403} {\bibfield  {journal} {\bibinfo
			{journal} {Phys. Rev. B}\ }\textbf {\bibinfo {volume} {109}},\ \bibinfo
		{pages} {184403} (\bibinfo {year} {2024})}\BibitemShut {NoStop}%
	\bibitem [{\citenamefont {Li}\ \emph {et~al.}(2016)\citenamefont {Li},
		\citenamefont {Adroja}, \citenamefont {Biswas}, \citenamefont {Baker},
		\citenamefont {Zhang}, \citenamefont {Liu}, \citenamefont {Tsirlin},
		\citenamefont {Gegenwart},\ and\ \citenamefont {Zhang}}]{Li097201}%
	\BibitemOpen
	\bibfield  {author} {\bibinfo {author} {\bibfnamefont {Y.}~\bibnamefont
			{Li}}, \bibinfo {author} {\bibfnamefont {D.}~\bibnamefont {Adroja}}, \bibinfo
		{author} {\bibfnamefont {P.~K.}\ \bibnamefont {Biswas}}, \bibinfo {author}
		{\bibfnamefont {P.~J.}\ \bibnamefont {Baker}}, \bibinfo {author}
		{\bibfnamefont {Q.}~\bibnamefont {Zhang}}, \bibinfo {author} {\bibfnamefont
			{J.}~\bibnamefont {Liu}}, \bibinfo {author} {\bibfnamefont {A.~A.}\
			\bibnamefont {Tsirlin}}, \bibinfo {author} {\bibfnamefont {P.}~\bibnamefont
			{Gegenwart}},\ and\ \bibinfo {author} {\bibfnamefont {Q.}~\bibnamefont
			{Zhang}},\ }\bibfield  {title} {\bibinfo {title} {{Muon Spin Relaxation
				Evidence for the U(1) Quantum Spin-Liquid Ground State in the Triangular
				Antiferromagnet ${\mathrm{YbMgGaO}}_{4}$}},\ }\href
	{https://doi.org/10.1103/PhysRevLett.117.097201} {\bibfield  {journal}
		{\bibinfo  {journal} {Phys. Rev. Lett.}\ }\textbf {\bibinfo {volume} {117}},\
		\bibinfo {pages} {097201} (\bibinfo {year} {2016})}\BibitemShut {NoStop}%
	\bibitem [{\citenamefont {Bhattacharya}\ \emph {et~al.}(2024)\citenamefont
		{Bhattacharya}, \citenamefont {Mohanty}, \citenamefont {Hillier},
		\citenamefont {Telling}, \citenamefont {Nath},\ and\ \citenamefont
		{Majumder}}]{BhattacharyaL060403}%
	\BibitemOpen
	\bibfield  {author} {\bibinfo {author} {\bibfnamefont {K.}~\bibnamefont
			{Bhattacharya}}, \bibinfo {author} {\bibfnamefont {S.}~\bibnamefont
			{Mohanty}}, \bibinfo {author} {\bibfnamefont {A.~D.}\ \bibnamefont
			{Hillier}}, \bibinfo {author} {\bibfnamefont {M.~T.~F.}\ \bibnamefont
			{Telling}}, \bibinfo {author} {\bibfnamefont {R.}~\bibnamefont {Nath}},\ and\
		\bibinfo {author} {\bibfnamefont {M.}~\bibnamefont {Majumder}},\ }\bibfield
	{title} {\bibinfo {title} {{Evidence of quantum spin liquid state in a
				${\mathrm{Cu}}^{2+}$-based $S=\frac{1}{2}$ triangular lattice
				antiferromagnet}},\ }\href {https://doi.org/10.1103/PhysRevB.110.L060403}
	{\bibfield  {journal} {\bibinfo  {journal} {Phys. Rev. B}\ }\textbf {\bibinfo
			{volume} {110}},\ \bibinfo {pages} {L060403} (\bibinfo {year}
		{2024})}\BibitemShut {NoStop}%
	\bibitem [{\citenamefont {Mendels}\ \emph {et~al.}(2007)\citenamefont
		{Mendels}, \citenamefont {Bert}, \citenamefont {de~Vries}, \citenamefont
		{Olariu}, \citenamefont {Harrison}, \citenamefont {Duc}, \citenamefont
		{Trombe}, \citenamefont {Lord}, \citenamefont {Amato},\ and\ \citenamefont
		{Baines}}]{Mendels077204}%
	\BibitemOpen
	\bibfield  {author} {\bibinfo {author} {\bibfnamefont {P.}~\bibnamefont
			{Mendels}}, \bibinfo {author} {\bibfnamefont {F.}~\bibnamefont {Bert}},
		\bibinfo {author} {\bibfnamefont {M.~A.}\ \bibnamefont {de~Vries}}, \bibinfo
		{author} {\bibfnamefont {A.}~\bibnamefont {Olariu}}, \bibinfo {author}
		{\bibfnamefont {A.}~\bibnamefont {Harrison}}, \bibinfo {author}
		{\bibfnamefont {F.}~\bibnamefont {Duc}}, \bibinfo {author} {\bibfnamefont
			{J.~C.}\ \bibnamefont {Trombe}}, \bibinfo {author} {\bibfnamefont {J.~S.}\
			\bibnamefont {Lord}}, \bibinfo {author} {\bibfnamefont {A.}~\bibnamefont
			{Amato}},\ and\ \bibinfo {author} {\bibfnamefont {C.}~\bibnamefont
			{Baines}},\ }\bibfield  {title} {\bibinfo {title} {{Quantum Magnetism in the
				Paratacamite Family: Towards an Ideal Kagom\'e Lattice}},\ }\href
	{https://doi.org/10.1103/PhysRevLett.98.077204} {\bibfield  {journal}
		{\bibinfo  {journal} {Phys. Rev. Lett.}\ }\textbf {\bibinfo {volume} {98}},\
		\bibinfo {pages} {077204} (\bibinfo {year} {2007})}\BibitemShut {NoStop}%
	\bibitem [{\citenamefont {Cho}\ \emph {et~al.}(2020)\citenamefont {Cho},
		\citenamefont {Nirmala}, \citenamefont {Jeong}, \citenamefont {Baker},
		\citenamefont {Takeda}, \citenamefont {Mera}, \citenamefont {Blundell},
		\citenamefont {Takigawa}, \citenamefont {Adroja},\ and\ \citenamefont
		{Park}}]{Cho014439}%
	\BibitemOpen
	\bibfield  {author} {\bibinfo {author} {\bibfnamefont {H.}~\bibnamefont
			{Cho}}, \bibinfo {author} {\bibfnamefont {R.}~\bibnamefont {Nirmala}},
		\bibinfo {author} {\bibfnamefont {J.}~\bibnamefont {Jeong}}, \bibinfo
		{author} {\bibfnamefont {P.~J.}\ \bibnamefont {Baker}}, \bibinfo {author}
		{\bibfnamefont {H.}~\bibnamefont {Takeda}}, \bibinfo {author} {\bibfnamefont
			{N.}~\bibnamefont {Mera}}, \bibinfo {author} {\bibfnamefont {S.~J.}\
			\bibnamefont {Blundell}}, \bibinfo {author} {\bibfnamefont {M.}~\bibnamefont
			{Takigawa}}, \bibinfo {author} {\bibfnamefont {D.~T.}\ \bibnamefont
			{Adroja}},\ and\ \bibinfo {author} {\bibfnamefont {J.-G.}\ \bibnamefont
			{Park}},\ }\bibfield  {title} {\bibinfo {title} {{Dynamic spin fluctuations
				in the frustrated $A$-site spinel
				$\mathrm{Cu}{\mathrm{Al}}_{2}{\mathrm{O}}_{4}$}},\ }\href
	{https://doi.org/10.1103/PhysRevB.102.014439} {\bibfield  {journal} {\bibinfo
			{journal} {Phys. Rev. B}\ }\textbf {\bibinfo {volume} {102}},\ \bibinfo
		{pages} {014439} (\bibinfo {year} {2020})}\BibitemShut {NoStop}%
	\bibitem [{\citenamefont {Hermele}\ \emph {et~al.}(2008)\citenamefont
		{Hermele}, \citenamefont {Ran}, \citenamefont {Lee},\ and\ \citenamefont
		{Wen}}]{Hermele224413}%
	\BibitemOpen
	\bibfield  {author} {\bibinfo {author} {\bibfnamefont {M.}~\bibnamefont
			{Hermele}}, \bibinfo {author} {\bibfnamefont {Y.}~\bibnamefont {Ran}},
		\bibinfo {author} {\bibfnamefont {P.~A.}\ \bibnamefont {Lee}},\ and\ \bibinfo
		{author} {\bibfnamefont {X.-G.}\ \bibnamefont {Wen}},\ }\bibfield  {title}
	{\bibinfo {title} {{Properties of an algebraic spin liquid on the kagome
				lattice}},\ }\href {https://doi.org/10.1103/PhysRevB.77.224413} {\bibfield
		{journal} {\bibinfo  {journal} {Phys. Rev. B}\ }\textbf {\bibinfo {volume}
			{77}},\ \bibinfo {pages} {224413} (\bibinfo {year} {2008})}\BibitemShut
	{NoStop}%
	\bibitem [{\citenamefont {Zhou}\ \emph {et~al.}(2008)\citenamefont {Zhou},
		\citenamefont {Lee}, \citenamefont {Ng},\ and\ \citenamefont
		{Zhang}}]{Zhou197201}%
	\BibitemOpen
	\bibfield  {author} {\bibinfo {author} {\bibfnamefont {Y.}~\bibnamefont
			{Zhou}}, \bibinfo {author} {\bibfnamefont {P.~A.}\ \bibnamefont {Lee}},
		\bibinfo {author} {\bibfnamefont {T.-K.}\ \bibnamefont {Ng}},\ and\ \bibinfo
		{author} {\bibfnamefont {F.-C.}\ \bibnamefont {Zhang}},\ }\bibfield  {title}
	{\bibinfo {title} {{${\mathrm{Na}}_{4}{\mathrm{Ir}}_{3}{\mathrm{O}}_{8}$ as a
				3D Spin Liquid with Fermionic Spinons}},\ }\href
	{https://doi.org/10.1103/PhysRevLett.101.197201} {\bibfield  {journal}
		{\bibinfo  {journal} {Phys. Rev. Lett.}\ }\textbf {\bibinfo {volume} {101}},\
		\bibinfo {pages} {197201} (\bibinfo {year} {2008})}\BibitemShut {NoStop}%
	\bibitem [{\citenamefont {Xu}\ \emph {et~al.}(2025)\citenamefont {Xu},
		\citenamefont {Du}, \citenamefont {Cheong},\ and\ \citenamefont
		{Cava}}]{Xu2400308}%
	\BibitemOpen
	\bibfield  {author} {\bibinfo {author} {\bibfnamefont {X.}~\bibnamefont
			{Xu}}, \bibinfo {author} {\bibfnamefont {K.}~\bibnamefont {Du}}, \bibinfo
		{author} {\bibfnamefont {S.-W.}\ \bibnamefont {Cheong}},\ and\ \bibinfo
		{author} {\bibfnamefont {R.~J.}\ \bibnamefont {Cava}},\ }\bibfield  {title}
	{\bibinfo {title} {Anomalous magnetoelectric coupling in the paramagnetic
			state of a chiral and polar magnet},\ }\href
	{https://doi.org/10.1002/aelm.202400308} {\bibfield  {journal} {\bibinfo
			{journal} {Adv. Electronic Mater.}\ }\textbf {\bibinfo {volume} {11}},\
		\bibinfo {pages} {2400308} (\bibinfo {year} {2025})}\BibitemShut {NoStop}%
	\bibitem [{\citenamefont {Sheldrick}(2018)}]{Sheldrick2018shelxl}%
	\BibitemOpen
	\bibfield  {author} {\bibinfo {author} {\bibfnamefont {G.~M.}\ \bibnamefont
			{Sheldrick}},\ }\bibfield  {title} {\bibinfo {title} {Shelxl-2018/3 software
			package},\ }\href@noop {} {\bibfield  {journal} {\bibinfo  {journal}
			{University of G{\"o}ttingen, Germany}\ } (\bibinfo {year}
		{2018})}\BibitemShut {NoStop}%
	\bibitem [{\citenamefont {Tsirlin}\ \emph {et~al.}(2024)\citenamefont
		{Tsirlin}, \citenamefont {Ginga},\ and\ \citenamefont {Martin}}]{esrf}%
	\BibitemOpen
	\bibfield  {author} {\bibinfo {author} {\bibfnamefont {A.~A.}\ \bibnamefont
			{Tsirlin}}, \bibinfo {author} {\bibfnamefont {V.~A.}\ \bibnamefont {Ginga}},\
		and\ \bibinfo {author} {\bibfnamefont {J.~G.}\ \bibnamefont {Martin}},\
	}\bibfield  {title} {\bibinfo {title} {High-resolution powder diffraction
			study of the double trillium lattice magnet {K$_2$Co$_2$(SO$_4)_3$}},\
	}\bibfield  {journal} {\bibinfo  {journal} {European Synchrotron Radiation
			Facility (ESRF)}\ }\href {https://doi.org/10.15151/ESRF-DC-2217270524}
	{10.15151/ESRF-DC-2217270524} (\bibinfo {year} {2024})\BibitemShut {NoStop}%
	\bibitem [{\citenamefont {Fitch}\ \emph {et~al.}(2023)\citenamefont {Fitch},
		\citenamefont {Dejoie}, \citenamefont {Covacci}, \citenamefont
		{Confalonieri}, \citenamefont {Grendal}, \citenamefont {Claustre},
		\citenamefont {Guillou}, \citenamefont {Kieffer}, \citenamefont {de~Nolf},
		\citenamefont {Petitdemange}, \citenamefont {Ruat},\ and\ \citenamefont
		{Watier}}]{Fitch1003}%
	\BibitemOpen
	\bibfield  {author} {\bibinfo {author} {\bibfnamefont {A.}~\bibnamefont
			{Fitch}}, \bibinfo {author} {\bibfnamefont {C.}~\bibnamefont {Dejoie}},
		\bibinfo {author} {\bibfnamefont {E.}~\bibnamefont {Covacci}}, \bibinfo
		{author} {\bibfnamefont {G.}~\bibnamefont {Confalonieri}}, \bibinfo {author}
		{\bibfnamefont {O.}~\bibnamefont {Grendal}}, \bibinfo {author} {\bibfnamefont
			{L.}~\bibnamefont {Claustre}}, \bibinfo {author} {\bibfnamefont
			{P.}~\bibnamefont {Guillou}}, \bibinfo {author} {\bibfnamefont
			{J.}~\bibnamefont {Kieffer}}, \bibinfo {author} {\bibfnamefont
			{W.}~\bibnamefont {de~Nolf}}, \bibinfo {author} {\bibfnamefont
			{S.}~\bibnamefont {Petitdemange}}, \bibinfo {author} {\bibfnamefont
			{M.}~\bibnamefont {Ruat}},\ and\ \bibinfo {author} {\bibfnamefont
			{Y.}~\bibnamefont {Watier}},\ }\bibfield  {title} {\bibinfo {title} {{ID22
				{--} the high-resolution powder-diffraction beamline at ESRF}},\ }\href
	{https://doi.org/10.1107/S1600577523004915} {\bibfield  {journal} {\bibinfo
			{journal} {J. Synch. Rad.}\ }\textbf {\bibinfo {volume} {30}},\ \bibinfo
		{pages} {1003} (\bibinfo {year} {2023})}\BibitemShut {NoStop}%
	\bibitem [{\citenamefont {Tsirlin}\ \emph {et~al.}(2009)\citenamefont
		{Tsirlin}, \citenamefont {Schmidt}, \citenamefont {Skourski}, \citenamefont
		{Nath}, \citenamefont {Geibel},\ and\ \citenamefont
		{Rosner}}]{Tsirlin132407}%
	\BibitemOpen
	\bibfield  {author} {\bibinfo {author} {\bibfnamefont {A.~A.}\ \bibnamefont
			{Tsirlin}}, \bibinfo {author} {\bibfnamefont {B.}~\bibnamefont {Schmidt}},
		\bibinfo {author} {\bibfnamefont {Y.}~\bibnamefont {Skourski}}, \bibinfo
		{author} {\bibfnamefont {R.}~\bibnamefont {Nath}}, \bibinfo {author}
		{\bibfnamefont {C.}~\bibnamefont {Geibel}},\ and\ \bibinfo {author}
		{\bibfnamefont {H.}~\bibnamefont {Rosner}},\ }\bibfield  {title} {\bibinfo
		{title} {{Exploring the spin-$\frac{1}{2}$ frustrated square lattice model
				with high-field magnetization studies}},\ }\href
	{https://doi.org/10.1103/PhysRevB.80.132407} {\bibfield  {journal} {\bibinfo
			{journal} {Phys. Rev. B}\ }\textbf {\bibinfo {volume} {80}},\ \bibinfo
		{pages} {132407} (\bibinfo {year} {2009})}\BibitemShut {NoStop}%
	\bibitem [{\citenamefont {Pratt}(2000)}]{Pratt710}%
	\BibitemOpen
	\bibfield  {author} {\bibinfo {author} {\bibfnamefont {F.}~\bibnamefont
			{Pratt}},\ }\bibfield  {title} {\bibinfo {title} {{WIMDA}: a muon data
			analysis program for the {Windows PC}},\ }\href
	{https://doi.org/https://doi.org/10.1016/S0921-4526(00)00328-8} {\bibfield
		{journal} {\bibinfo  {journal} {Physica B Condens. Matter}\ }\textbf
		{\bibinfo {volume} {289-290}},\ \bibinfo {pages} {710} (\bibinfo {year}
		{2000})}\BibitemShut {NoStop}%
	\bibitem [{\citenamefont {Kresse}\ and\ \citenamefont
		{Furthm\"uller}(1996{\natexlab{a}})}]{vasp1}%
	\BibitemOpen
	\bibfield  {author} {\bibinfo {author} {\bibfnamefont {G.}~\bibnamefont
			{Kresse}}\ and\ \bibinfo {author} {\bibfnamefont {J.}~\bibnamefont
			{Furthm\"uller}},\ }\bibfield  {title} {\bibinfo {title} {Efficiency of
			\textit{ab-initio} total energy calculations for metals and semiconductors
			using a plane-wave basis set},\ }\href
	{https://doi.org/10.1016/0927-0256(96)00008-0} {\bibfield  {journal}
		{\bibinfo  {journal} {Comput. Mater. Sci.}\ }\textbf {\bibinfo {volume}
			{6}},\ \bibinfo {pages} {15} (\bibinfo {year}
		{1996}{\natexlab{a}})}\BibitemShut {NoStop}%
	\bibitem [{\citenamefont {Kresse}\ and\ \citenamefont
		{Furthm\"uller}(1996{\natexlab{b}})}]{vasp2}%
	\BibitemOpen
	\bibfield  {author} {\bibinfo {author} {\bibfnamefont {G.}~\bibnamefont
			{Kresse}}\ and\ \bibinfo {author} {\bibfnamefont {J.}~\bibnamefont
			{Furthm\"uller}},\ }\bibfield  {title} {\bibinfo {title} {Efficient iterative
			schemes for \textit{ab initio} total-energy calculations using a plane-wave
			basis set},\ }\href {https://doi.org/10.1103/PhysRevB.54.11169} {\bibfield
		{journal} {\bibinfo  {journal} {Phys. Rev. B}\ }\textbf {\bibinfo {volume}
			{54}},\ \bibinfo {pages} {11169} (\bibinfo {year}
		{1996}{\natexlab{b}})}\BibitemShut {NoStop}%
	\bibitem [{\citenamefont {Perdew}\ \emph {et~al.}(1996)\citenamefont {Perdew},
		\citenamefont {Burke},\ and\ \citenamefont {Ernzerhof}}]{pbe96}%
	\BibitemOpen
	\bibfield  {author} {\bibinfo {author} {\bibfnamefont {J.~P.}\ \bibnamefont
			{Perdew}}, \bibinfo {author} {\bibfnamefont {K.}~\bibnamefont {Burke}},\ and\
		\bibinfo {author} {\bibfnamefont {M.}~\bibnamefont {Ernzerhof}},\ }\bibfield
	{title} {\bibinfo {title} {Generalized gradient approximation made simple},\
	}\href {https://doi.org/10.1103/PhysRevLett.77.3865} {\bibfield  {journal}
		{\bibinfo  {journal} {Phys. Rev. Lett.}\ }\textbf {\bibinfo {volume} {77}},\
		\bibinfo {pages} {3865} (\bibinfo {year} {1996})}\BibitemShut {NoStop}%
	\bibitem [{\citenamefont {Tsirlin}(2014)}]{tsirlin2014}%
	\BibitemOpen
	\bibfield  {author} {\bibinfo {author} {\bibfnamefont {A.~A.}\ \bibnamefont
			{Tsirlin}},\ }\bibfield  {title} {\bibinfo {title} {Spin-chain magnetism and
			uniform {Dzyaloshinsky-Moriya} anisotropy in {BaV$_3$O$_8$}},\ }\href
	{https://doi.org/10.1103/PhysRevB.89.014405} {\bibfield  {journal} {\bibinfo
			{journal} {Phys. Rev. B}\ }\textbf {\bibinfo {volume} {89}},\ \bibinfo
		{pages} {014405} (\bibinfo {year} {2014})}\BibitemShut {NoStop}%
	\bibitem [{\citenamefont {Bader}\ \emph {et~al.}(2022)\citenamefont {Bader},
		\citenamefont {Langmann}, \citenamefont {Gegenwart},\ and\ \citenamefont
		{Tsirlin}}]{bader2022}%
	\BibitemOpen
	\bibfield  {author} {\bibinfo {author} {\bibfnamefont {V.~P.}\ \bibnamefont
			{Bader}}, \bibinfo {author} {\bibfnamefont {J.}~\bibnamefont {Langmann}},
		\bibinfo {author} {\bibfnamefont {P.}~\bibnamefont {Gegenwart}},\ and\
		\bibinfo {author} {\bibfnamefont {A.~A.}\ \bibnamefont {Tsirlin}},\
	}\bibfield  {title} {\bibinfo {title} {Deformation of the triangular
			spin-$\frac12$ lattice in {Na$_2$SrCo(PO$_4)_2$}},\ }\href
	{https://doi.org/10.1103/PhysRevB.106.054415} {\bibfield  {journal} {\bibinfo
			{journal} {Phys. Rev. B}\ }\textbf {\bibinfo {volume} {106}},\ \bibinfo
		{pages} {054415} (\bibinfo {year} {2022})}\BibitemShut {NoStop}%
	\bibitem [{\citenamefont {Solana-Madruga}\ \emph {et~al.}(2024)\citenamefont
		{Solana-Madruga}, \citenamefont {Mentr\'e}, \citenamefont {Tsirlin},
		\citenamefont {Huv\'e}, \citenamefont {Khalyavin}, \citenamefont {Ritter},\
		and\ \citenamefont {Ar\'evalo-L\'opez}}]{madruga2024}%
	\BibitemOpen
	\bibfield  {author} {\bibinfo {author} {\bibfnamefont {E.}~\bibnamefont
			{Solana-Madruga}}, \bibinfo {author} {\bibfnamefont {O.}~\bibnamefont
			{Mentr\'e}}, \bibinfo {author} {\bibfnamefont {A.~A.}\ \bibnamefont
			{Tsirlin}}, \bibinfo {author} {\bibfnamefont {M.}~\bibnamefont {Huv\'e}},
		\bibinfo {author} {\bibfnamefont {D.}~\bibnamefont {Khalyavin}}, \bibinfo
		{author} {\bibfnamefont {C.}~\bibnamefont {Ritter}},\ and\ \bibinfo {author}
		{\bibfnamefont {A.~M.}\ \bibnamefont {Ar\'evalo-L\'opez}},\ }\bibfield
	{title} {\bibinfo {title} {{CoVO$_3$} high-pressure polymorphs: To order or
			not to order?},\ }\href {https://doi.org/10.1002/advs.202307766} {\bibfield
		{journal} {\bibinfo  {journal} {Adv. Science}\ }\textbf {\bibinfo {volume}
			{11}},\ \bibinfo {pages} {2307766} (\bibinfo {year} {2024})}\BibitemShut
	{NoStop}%
	\bibitem [{\citenamefont {Xu}\ \emph {et~al.}(2016)\citenamefont {Xu},
		\citenamefont {Zhang}, \citenamefont {Li}, \citenamefont {Yu}, \citenamefont
		{Hong}, \citenamefont {Zhang},\ and\ \citenamefont {Li}}]{Xu267202}%
	\BibitemOpen
	\bibfield  {author} {\bibinfo {author} {\bibfnamefont {Y.}~\bibnamefont
			{Xu}}, \bibinfo {author} {\bibfnamefont {J.}~\bibnamefont {Zhang}}, \bibinfo
		{author} {\bibfnamefont {Y.~S.}\ \bibnamefont {Li}}, \bibinfo {author}
		{\bibfnamefont {Y.~J.}\ \bibnamefont {Yu}}, \bibinfo {author} {\bibfnamefont
			{X.~C.}\ \bibnamefont {Hong}}, \bibinfo {author} {\bibfnamefont {Q.~M.}\
			\bibnamefont {Zhang}},\ and\ \bibinfo {author} {\bibfnamefont {S.~Y.}\
			\bibnamefont {Li}},\ }\bibfield  {title} {\bibinfo {title} {{Absence of
				Magnetic Thermal Conductivity in the Quantum Spin-Liquid Candidate
				${\mathrm{YbMgGaO}}_{4}$}},\ }\href
	{https://doi.org/10.1103/PhysRevLett.117.267202} {\bibfield  {journal}
		{\bibinfo  {journal} {Phys. Rev. Lett.}\ }\textbf {\bibinfo {volume} {117}},\
		\bibinfo {pages} {267202} (\bibinfo {year} {2016})}\BibitemShut {NoStop}%
\end{thebibliography}

%

\end{document}